\documentclass[11 pt,nofootinbib,notitlepage,eqsecnum]{revtex4-1}
\usepackage[english]{babel}
\usepackage{bbm}
\usepackage{mathtools} 
\usepackage{soul,xcolor}
\setstcolor{red} 

\showboxdepth=\maxdimen
\showboxbreadth=\maxdimen

\usepackage{subfig}
\usepackage{graphicx,tikz}
\usetikzlibrary{arrows,shapes.geometric,positioning,shapes.misc}
\usepgfmodule{shapes}
\usetikzlibrary{decorations.markings,decorations.text,decorations,patterns,calc}
\usepackage{pgfkeys}

\renewcommand{\bm}[1]{{\mathbf #1}}

\usepackage{amsmath,amssymb,amsthm,MnSymbol}

\usepackage{hyperref}

\newtheorem{lemma}{Lemma}
\newtheorem{corollary}{Corollary}
\newtheorem{theorem}{Theorem}
\newtheorem{definition}{Definition}

\newcommand{\startproof}{\noindent\textbf{Proof.} }
\newcommand{\finishproof}{\hfill $\blacksquare$ \\\hspace*{\fill}}

\newcommand{\dif}{\mathrm{d}}

\newcommand{\eqat}[1]{\mathrel{\text{\setbox0\hbox{$=$}\rlap{\hbox to \wd0{\hss\raisebox{-.70\height}{$\scriptscriptstyle #1$}\hss}}\box0}}}

\newcommand{\ee}[1]{\operatorname{e}^{#1}}

\def\Re{\mathrm{Re}\,}
\def\Im{\mathrm{Im}\,}
\def\tr{\mathrm{tr}}
\def\half{\frac{1}{2}}
\def\sgn{\mathrm{sgn}}

\newcommand{\ket}[1]{|#1\rangle}
\newcommand{\bra}[1]{\langle #1|}
\newcommand{\scal}[1]{\langle #1\rangle}

\DeclarePairedDelimiter{\abs}{\lvert}{\rvert}

\def\R{\mathbb{R}}

\def\C{\mathbb{C}}
\def\CP{\mathbb{CP}}

\def\timeo{\mathcal{T}}

\newcommand{\SL}{\mathrm{SL}(2,\C)}

\newcommand{\SU}{\mathrm{SU}}

\newcommand{\pauli}{{\boldsymbol\sigma}}

\newcommand{\boost}{{\bm K}}
\def\rotoperator{\hat{\,\bm L}}

\newcommand{\Crit}{\mathrm{Crit}}

\def\SFreview{\cite{Baez:1999sr,Perez:2004hj,Perez:2012wv,Rovelli:2011eq}}
\def\LQGreview{\cite{ThiemannBook, Ashtekar:2004eh,RovelliBook,Sahlmann:2010zf}}

\def\EPRLdev{\cite{Engle:2007uq,Engle:2007qf,Freidel:2007py,Pereira:2007nh,Engle:2007wy, Engle:2008ev}}

\def\asymptoticL{\cite{Barrett:2009mw}}
\def\asymptoticE{\cite{Barrett:2009gg}}

\def\jonProp{\cite{Engle:2011un}}

\def\beq{\begin{equation}}
\def\eeq{\end{equation}}
\def\bq{\begin{equation*}}
\def\eq{\end{equation*}}

\newcommand{\propdec}{\mathrm{(+)}}
\newcommand{\Sprop}{S^{\propdec}}
\newcommand{\Aprop}{A^{\propdec}_v}
\newcommand{\Apropi}[1]{A^\propdec_{#1}}

\newcommand{\x}[1]{x_{#1}} 
\newcommand{\y}[1]{y_{#1}} 
\newcommand{\A}[1]{A^{#1}} 
\newcommand{\vbasis}[2]{(v^{#1})_{#2}} 
\newcommand{\ucoord}[1]{u_{#1}}
\newcommand{\covdelta}[1]{\delta^{#1}} 

\begin{document}
\title{The Lorentzian proper vertex amplitude: Asymptotics}

\author{Jonathan Engle${}^{1,2}$}
\email{jonathan.engle@fau.edu}

\author{Ilya Vilensky${}^{1}$}
\email{ilya.vilensky@fau.edu}

\author{Antonia Zipfel${}^{2,3}$}
\email{antonia.zipfel@fuw.pl}

\affiliation{${}^1$ \mbox{Department of Physics, Florida Atlantic University, Boca Raton, Florida, USA}\\
${}^2$ \mbox{Institut f\"ur Quantengravitation, Department Physik Universit\"at Erlangen,}\\ Staudtstrasse 7, D-91058 Erlangen, Germany\\
${}^3$ \mbox{Instytut Fizyki Teoretycznej, Uniwersytet Warszawski, ul. Ho\.za 69, 00-681 Warsaw, Poland}}

\begin{abstract}
\centerline{\bf Abstract}

In previous work, the Lorentzian proper vertex amplitude for a spin-foam model of quantum gravity was derived.
In the present work, the asymptotics of this amplitude are studied in the semi-classical limit.
The starting point of the analysis is an expression for the amplitude as an action integral with
action differing from that in the EPRL case by an extra `projector' term which scales linearly with spins
only in the asymptotic limit, and is discontinuous on a submanifold of the integration domain.
New tools are introduced to generalize stationary phase methods to this case.
For the case of boundary data which can be glued to a non-degenerate Lorentzian 4-simplex,
the asymptotic limit of the amplitude is shown to equal the single Feynman term,
showing that the extra term in the asymptotics of the EPRL amplitude has been eliminated.
\end{abstract}
\maketitle

\section{Introduction}

Spin foam models {\SFreview} arise from a covariant formulation of loop quantum gravity {\LQGreview}. These models are usually developed on a triangulated manifold and defined by an amplitude associated with the simplices of the triangulation.
A crucial tool in understanding the semi-classical regime of a spin-foam amplitude is the analysis of its asymptotic behavior. When Ponzano and Regge discovered that the asymptotics of the 6j-symbol in recoupling theory contain the discrete Regge action for GR \cite{pr1968}, they established the first quantum model for Euclidean 3D gravity. The Ponzano-Regge model was generalized in spin-foam models of 4D gravity.  In the papers {\EPRLdev} were developed Euclidean and Lorentzian versions of
what came to be known as the EPRL model.

The asymptotics of the Euclidean EPRL model were studied in {\asymptoticE}. It was found that in the semi-classical regime the EPRL amplitude contains several terms made of exponentials of the Regge action with different overall coefficients and different signs, instead of one exponential term. When dealing with multiple 4-simplices these extra terms give rise to unphysical equations of motion and perhaps are a cause of bubble divergences \cite{clrrr2012}.
The presence of similar problematic terms was discovered for the Lorentzian EPRL model in {\asymptoticL}. It was shown in
\cite{engle2011, ez2015} that the appearance of these terms is caused by the presence of  different
combinations of dynamical orientations and Plebanski sectors of solutions to the simplicity constraint.
The combination of Plebanski sectors and orientations yielding the one Feynman-like term
equal to the exponential of exactly $i$ times the Regge action was named the
\textit{Einstein-Hilbert sector}.
This name arises from the fact that,
in continuum Plebanski theory this is the sector in which the action reduces to
Einstein-Hilbert.
These observations led to the proposal in
\cite{Engle:2011un, Engle:2012yg} to quantize the restriction to the Einstein-Hilbert sector
by inserting an appropriate projector into the amplitude, thereby eliminating the additional terms.

In \cite{ez2015} this strategy was implemented to modify the Lorentzian EPRL amplitude and define what is called the
\textit{Lorentzian proper vertex amplitude}. In this paper we study the asymptotics of the Lorentzian proper vertex amplitude and verify that the semi-classical regime of the new amplitude is indeed dominated by a single term, the exponential of the Regge action. The analysis is performed by writing the amplitude in a path-integral form, splitting the action into a modified EPRL part and a projector action part,
and using stationary phase methods to study the large-spin limit.
However, the non-linear dependence of the projector action on spins as well as discontinuities in this term where the parallel transports are degenerate,
significantly complicates the analysis when one is careful.
As a consequence new strategies are used in this paper to extend the stationary phase methods to the case of non-linear actions
as well as to handle the points of discontinuity.
The present analysis restricts consideration to the case of boundary states for which there
are no degenerate critical points.  Thus the present analysis includes, in particular, 
the most physically relevant case of boundary data, 
namely that of those which fit onto a non-degenerate Lorentzian 4-simplex.

Note this paper concerns only verifying the correct \textit{phase} of the asymptotics of the proper vertex.
This is the most important part of the asymptotics to check as it is necessary to ensure that the classical equations of motion
dominate the spin-foam sum in the semi-classical limit. The other part of the asymptotics --- the modulus ---
determines the effective measure factor of the path integral, which is important for determining formal equivalence with
canonical quantization \cite{ce2013, Engle:2009ba}.

The paper is organized as follows.  A brief review of the key equations and prior results needed for the paper are given in section \ref{sec:prelim}.
In section \ref{sec:propasym} we formulate the amplitude in integral form, consider the critical points of the action and calculate their contribution to asymptotics, which is the main result of the paper. In section \ref{sec:completion}, we present the technical argument justifying the use of stationary phase methods for evaluating the asymptotics. In particular, we derive the asymptotic form of the non-linear part of the action,
show that the usual stationary phase theorem applies to the part of the vertex integral containing all critical points,
and show that the contribution from the rest of the vertex integral is suppressed.
\newpage

\section{Preliminaries}
\label{sec:prelim}

\subsection{EPRL asymptotics}
\label{eprl_asympt}

To define a vertex amplitude we consider a single 4-simplex with tetrahedra labelled $a, b = 0 \dots 4 $ and triangles labelled by unordered pairs $(ab)$.
The boundary Hilbert space is spanned by generalized spin-network states,
in which each triangle is labelled by a spin $k_{ab}$ and two vectors $\psi_{ab}, \psi_{ba}$ in the corresponding irreducible representation of $\SU(2)$.
For the purpose of this paper, we use coherent boundary states consisting in spin-network states with the vectors  $\psi_{ab}, \psi_{ba}$ chosen to be Perelomov coherent states (\cite{PerelomovGCS}) associated to two unit spinors $\xi_{ab}, \xi_{ba}$ (for more details see \cite{ez2015,Barrett:2009mw}).
Each two-spinor $\xi$ determines a unit $3$-vector via \cite{ez2015,Barrett:2009mw}
\begin{align}
\label{eqn:ndef}
\mathbf{n}_\xi := \frac{\langle \xi | \pauli | \xi \rangle}{\langle \xi, \xi \rangle},
\end{align}
where $\pauli^i$ are the usual Pauli matrices.
Both the set $(k_{ab}, \xi_{ab})$ and the set $(k_{ab}, \mathbf{n}_{ab})$ which they determined, are called \textit{boundary data} for the 4-simplex.  Given $(k_{ab}, \mathbf{n}_{ab})$, $(k_{ab}, \xi_{ab})$
is uniquely determined up to a phase for each spinor.
The boundary data $(k_{ab}, \mathbf{n}_{ab})$ determine a geometry for each tetrahedron \cite{ez2015},
and are called \textit{Regge-like} when these geometries glue
together consistently to define a boundary geometry of a 4-simplex. In this case one can fix the phase of the boundary state leading to what is called a
\textit{Regge state}, which we denote $\psi^{\text{Regge}}_{k_{ab},\textbf{n}_{ab}}$.
The boundary data $(k_{ab}, \mathbf{n}_{ab})$ is called a \textit{vector geometry} when
the determined tetrahedra can be rotated in $\R^3$
such that triangle $(ab)$ in tetrahedron $a$ is parallel to triangle $(ab)$ in tetrahedron $b$, with opposite orientation, for all $a,b$
{\asymptoticE}.

In {\asymptoticL}, the asymptotics of the Lorentzian EPRL vertex amplitude were analyzed and found to be governed by critical points of an action. The amplitude of a Regge-state at a critical point depends only on $k_{ab}$ and the dihedral angle $\Theta_{ab}$ for which $\mathrm{cosh}\,\Theta_{ab}=|N_a\cdot N_b|$ if the boundary glues to a Lorentzian 4-simplex
with tetrahedron normals $N_a$, or $\cos\Theta^E_{ab}=|N^E_a\cdot N^E_b|$ if it glues to a Euclidean 4-simplex with
normals $N_a^E$.
We quote the result from {\asymptoticL}:
\begin{theorem}[EPRL asymptotics]
\label{asym_thm}
Let $\mathcal{B}=\{\lambda k_{ab},\bm{n}_{ab}\}$
%
%
be a set of non-degenerate boundary data satisfying closure.
\begin{enumerate}
\item
If $\mathcal{B}$ is Regge-like and determines the boundary geometry of a Lorentzian 4-simplex, then in the limit $\lambda \rightarrow \infty$,
%
%
\begin{align}
\label{eqn:asymL}
 A_v(\psi^{\rm Regge}_{\mathcal{B}})\sim\left(\frac{1}{\lambda}\right)^{12}\left[N_+ \exp\left(i\lambda\gamma \sum_{a<b} k_{ab}\Theta_{ab}\right)
+
N_- \exp\left(- i \lambda\gamma\sum_{a<b} k_{ab} \Theta_{ab}\right)
\right]~.
\end{align}
\item
If $\mathcal{B}$ is Regge-like and determines the boundary geometry of an Euclidean 4-simplex, then in the limit $\lambda \rightarrow \infty$,
\begin{align}
\label{eqn:asymE}
 A_v(\psi^{\rm Regge}_{\mathcal{B}})\sim\left(\frac{1}{\lambda}\right)^{12}\left[N_+^E \exp\left(i\lambda \sum_{a<b} k_{ab}\Theta^E_{ab}\right)
+
N_-^E \exp\left(- i \lambda\sum_{a<b} k_{ab} \Theta^E_{ab}\right)
\right]~.
\end{align}
\item
If $\mathcal{B}$ forms a vector geometry not in the above cases, then
\begin{align}
\label{vec_asym}
A_v \sim \left(\frac{2\pi}{\lambda}\right)^{12} N
\end{align}
 in the limit $\lambda \rightarrow \infty$.
\item
If $\mathcal{B}$ is not a vector geometry, then
$A_v$ falls off faster than any inverse power of $\lambda$.
\end{enumerate}
The factors $N_+$, $N_-$, $N_+^E$, $N_-^E$ and $N$  are independent of $\lambda$ and given in {\asymptoticL}
\end{theorem}
As shown in \cite{ez2015}, of all the terms above only the first term in (\ref{eqn:asymL})
arises from critical points corresponding to the Einstein-Hilbert sector.

\subsection{Proper vertex amplitude}

As defined in \cite{ez2015}, the proper EPRL-vertex for boundary data $\{k_{ab}, \xi_{ab}\}$ is given by
\begin{align}
\label{proper_vertex}
\Aprop(\{k_{ab}, \xi_{ab}\}):=(-1)^{\Xi}\int_{\SL^5}\delta(X_0)\prod_a\dif X_a\prod_{a<b}
\alpha(X_a\mathcal{I}C_{\xi_{ab}}^{k_{ab}},X_b\mathcal{I}\,\Pi_{ba}\left(\{X_{a'b'}\}\right)\,C_{\xi_{ba}}^{k_{ab}})~.
\end{align}
Here, $C_{\xi_{ab}}^{k_{ab}}$ is the coherent state in the spin $k_{ab}$ representation, associated with spinor $\xi_{ab}$ as described above
and defined in \cite{ez2015}.  $\mathcal{I}$ is the `EPRL' mapping from the spin $k_{ab}$ representation of $SU(2)$ into
the unitary irreducible representation of $\SL$ labeled by the $\SL$ Casimir quantum numbers $(k_{ab}, \gamma k_{ab})$
(see \cite{ez2015, Barrett:2009mw, Engle:2007wy}).  It is in the mapping $\mathcal{I}$ that the simplicity constraints
of the spin-foam model are encoded. $\alpha(\, ,\,)$ denotes the $\SL$-invariant bilinear form on the $(k_{ab}, \gamma k_{ab})$
representation of $\SL$ \cite{Barrett:2009mw, ez2015}.
The sign factor $(-1)^\Xi$ depends on the order of the tetrahedra and can be evaluated by a graphical calculus (see {\asymptoticL}).
Finally, $X_{ab}:= X_a^{-1} X_b$, and
\begin{align}
\label{eqn:quantum_projector}
\Pi_{ab}(\{X_{a'b'}\}):=\Pi_{(0,\infty)}\left(\beta_{ab}(\{X_{a'b'}\})\;(\widehat{X}_{ab} \timeo)_i\,
\rotoperator^i \right)
\end{align}
where $(\widehat{X}_{ab}\timeo)^i:= \half \tr(\pauli_i\, X_{ab}^{\mathstrut} \,X_{ab}^{\dagger})$
is the spatial part of the 4-vector $\timeo:= (1,0,0,0)$ rotated by the
$SO(3,1)$ action of $X_{ab}$,
$\rotoperator^i$ are the generators of $\SU(2)$,
$\Pi_{\cal S}(\hat{O})$ denotes the spectral projector onto the part ${\cal S}\subset\R$ of the spectrum of the operator $\hat{O}$, and where
\begin{align*}
\beta_{ab}(\{X_{a'b'}\}):= \mathrm{sgn}\left[\epsilon_{ijk}(\widehat{X}_{ac}\timeo)^i(\widehat{X}_{ad}\timeo)^j(\widehat{X}_{ae}\timeo)^k\; \epsilon_{lmn}(\widehat{X}_{bc}\timeo)^l(\widehat{X}_{bd}\timeo)^m(\widehat{X}_{be}\timeo)^n\right] .
\end{align*}

\section{Asymptotics}
\label{sec:propasym}

The asymptotic analysis of the proper EPRL-vertex is carried out in 
almost the same manner as in {\asymptoticL}:
Using coherent boundary states,
the amplitude is rewritten in terms of the exponential of an action and then the critical points are calculated.
However, two additional subtleties appear:
(1.) The scaling behavior under $k_{ab} \mapsto \lambda k_{ab}$ of the new term in the action is not linear in $\lambda$,
but only asymptotically linear, and (2.) the projector \eqref{eqn:quantum_projector} does not depend on the norm of the 3-vector defined by $(\widehat{X}_{ab}\timeo)^i$. The projector is, therefore, discontinuous on the subset where one of these vectors is zero, that is, where $X_{ab}\in\SU(2)$ for some $a<b$.
These two subtleties prevent the usual extended stationary phase theorem \cite{hormander1983}
from applying directly.
For the present section we ignore these,
calculating the critical points and asymptotics in the usual way.
Then, in the next section, we will prove in detail that the above two subtleties do not affect the results.

\subsection{Integral representation}

From (\ref{proper_vertex}), the proper amplitude for a coherent boundary state with data $\{k_{ab}, \xi_{ab}\}$ is given by
\begin{align}
\label{cohproper}
\Aprop(\{k_{ab}, \psi_{ab}\}):=(-1)^{\Xi}\int_{\SL^5}\delta(X_0)\prod_a\dif X_a\prod_{a<b}
\alpha(X_a\mathcal{I}C_{ab},X_b\mathcal{I}\,\Pi_{ba}\left(\{X_{a'b'}\}\right)\,C_{ba})
\end{align}
where $C_{ab}:= C_{\xi_{ab}}^{k_{ab}}$.
The main idea is to separate the amplitude \eqref{cohproper} into a part depending on the original action $S^{EPRL}$ for which the asymptotics is known and a part depending on the newly introduced projector. This can be achieved by inserting a resolution of the identity in terms of the integral kernel
\begin{align}
\label{resolution}
K(z,z'):=\frac{d_k}{\pi}\int_{\CP^1}\Omega_{\hat{\eta}} \; C^k_{\hat{\eta}}(z)\; \overline{C^k_{\hat{\eta}}(z')}~,
\end{align}
into the amplitude \eqref{cohproper}, where
$d_k:= 2k+1$,
$\Omega_\eta:= \frac{i}{2}(\eta_0 d\eta_1 - \eta_1 d\eta_0)\wedge(\overline{\eta}_0 d\overline{\eta}_1 - \overline{\eta}_1 d\overline{\eta}_0)$
is the standard invariant 2-form on $\CP^1$ --- the
space of equivalence classes $[\eta]$ of non-zero two-spinors $\eta$ modulo rescaling by a complex number ---
and
$\hat{\eta}:= \eta/||\eta||$ so that $\Omega_{\hat{\eta}} = \Omega_\eta/||\eta||^4$.
(See appendix A in \cite{ez2015} for details.)
This yields
\bq
\Aprop=(-1)^{\Xi}\!\!\!\!\int\limits_{\SL^5}\!\!\!{\cal D}X\!\!\!\!\int\limits_{(\CP^1)^{10}}
\prod_{a<b}\,\frac{d_{k_{ab}}}{\pi}\;\Omega_{\hat{\eta}_{ba}}\;
\alpha\left(X_a\,\mathcal{I}\,C^{k_{ab}}_{\xi_{ab}},X_b\,\mathcal{I}\,C^{k_{ab}}_{\hat{\eta}_{ba}}\right)
\left(C^{k_{ab}}_{\hat{\eta}_{ba}},\Pi_{ba}\left(\{X_{a'b'}\}\right)\,C^{k_{ab}}_{\xi_{ba}}\right)
\eq
where ${\cal D} X=\delta(X_0)\prod_a\dif X_a$ and $(,)$ denotes the Hermitian inner product on the irreducible representation of $SU(2)$ to which its arguments belong. The (anti-)symmetric inner product $\alpha$ can be expanded using equations (15) in {\asymptoticL}, (3.8) and (3.9) in \cite{ez2015} so that the amplitude can be replaced by an exponential expression
\begin{align}
\label{eqn:newExP}
\Aprop
= (-1)^{\Xi}\,\,c \!\! \int\limits_{\SL^5}\!\!{\cal D}X\!\!\!
 \int\limits_{(\CP^1)^{20}}
\left(\prod_{a<b} \, \frac{d^2_{k_{ab}}}{\pi}\;\Omega_{ab} \,\Omega_{\hat{\eta}_{ba}}\right) \;\exp \Sprop
\end{align}
where $c:= \frac{(1+\gamma^2)^5}{[\pi(1-i\gamma)]^{10}}$,
$\Omega_{ab}:= \frac{\Omega_{z_{ab}}}{\langle Z_{ab}, Z_{ab}\rangle \langle Z_{ba}, Z_{ba}\rangle}$, and
\begin{align}
\label{eqn:propaction}
\Sprop[\{k_{ab},\xi_{ab}\}; \{X_a, z_{ab}, \eta_{ba}\}]
&:= S^{EPRL}[\{k_{ab}, \xi_{ab}, \eta_{ba}\}; \{X_a, z_{ab}\}] +
S^\Pi[\{k_{ab}, \xi_{ba}\}; \{X_a, \eta_{ba}\}] \\
&:= \sum\limits_{a<b} S^{EPRL}_{ab} + \sum\limits_{a<b} S^\Pi_{ab}
\end{align}
with
\begin{align}
\label{eqn:EPRL_ab}
S^{EPRL}_{ab}[\{X_{a'}\},z_{ab},\eta_{ba}]&:= k_{ab}\, \log \frac{\scal{Z_{ab},\xi_{ab}}^2
\scal{J\hat{\eta}_{ba},Z_{ba}}^2}{\scal{Z_{ab},Z_{ab}}\scal{Z_{ba},Z_{ba}}}
+i\gamma k_{ab}\,\log \frac{\scal{Z_{ba},Z_{ba}}}{\scal{Z_{ab},Z_{ab}}}, \\
\label{eqn:C_ab}
S^\Pi_{ab}[\{X_{a'}\},\eta_{ba}]&:= \log \left(C_{\hat{\eta}_{ba}}^{k_{ab}},
\Pi_{ba}\left(\{X_{a'b'}\}\right)\,C_{\xi_{ba}}^{k_{ab}}\right),
\end{align}
%
%
$Z_{ab}:= X_a^{-1} z_{ab}$, $Z_{ba}:=X_b^{-1} z_{ab}$,
$\langle \alpha, \beta \rangle := \overline{\alpha}_0 \beta_0 + \overline{\alpha}_1 \beta_1$
is the Hermitian inner product on $\C^2$,
and where $J$ denotes the anti-linear structure map
\begin{align*}
J: \left(\begin{array}{c}  \xi_0 \\ \xi_1 \end{array} \right) \mapsto \left(\begin{array}{c}  -\overline{\xi}_1 \\ \overline{\xi}_0 \end{array} \right).
\end{align*}
Note the definitions of $Z_{ab}$ and $Z_{ba}$ here differ from those in {\asymptoticL}
by $X_a \mapsto (X_a^{-1})^\dagger$, so that the expression in (\ref{eqn:EPRL_ab})
also differs from that in {\asymptoticL} by the same replacement.
This difference is due both to a change in convention regarding the way the group elements
$X_a$ are used as well as a change in convention concerning the family of coherent states used.
The motivation and need for this change of convention is explained in appendix C of \cite{ez2015}.

The proper action $\Sprop$
depends on three sets of independent variables, $\{X_a\}$, $\{z_{ab}\}_{a<b}$ and $\{\eta_{ba}\}_{a<b}$.
However, $S^\Pi$ does not scale linearly under $k_{ab} \mapsto \lambda k_{ab}$, but only
asymptotically linearly, and it is discontinuous everywhere
$X_{ab} \in SU(2)$ for some $a<b$, as already noted above.
Later, in section \ref{sec:completion}, we will show that, in spite of these subtleties, the conclusion of the stationary phase theorem
still applies, thereby justifying the rest of the calculations and conclusions in this section.

\subsection{Critical points in the non-degenerate sector}
\label{sec:critical_points}

The symmetries of the action $\Sprop$ are
(1.) for each $Y$ in $\SL$, a global Lorentz symmetry $X_a \mapsto YX_a$, $z_{ab} \mapsto (Y^\dagger)^{-1} z_{ab}$, and
(2.) for each vertex $a$, a spin lift symmetry sending $X_a \mapsto -X_a$ and leaving all other variables
fixed.\footnote{In
addition to these, for each triangle $a<b$ and each non-zero $\kappa \in \C$,
the action is invariant under each of the rescalings
$z_{ab} \mapsto \kappa z_{ab}$ and $\eta_{ba} \mapsto \kappa \eta_{ba}$.
However, strictly speaking,
these are not symmetries but necessary conditions for the action to be a well-defined function of each
$[z_{ab}], [\eta_{ba}] \in \CP^1$.
}
Recall the critical points of an action $S$ are the stationary points for which ${\rm Re} \,S$ is maximal.
The symmetries of the action obviously map critical points to critical points.
Any two critical points related by such a symmetry we call \textit{equivalent}
because they will yield the same contribution to the asymptotics. In the following we will also assume that $X_{ab}\nin\SU(2)$ so that the projector is well defined.

\subsubsection{Maximality and stationarity with respect to \texorpdfstring{$z$}{z}}
\label{sec:zcrit}

From the triangle inequality follows immediately that
\bq
\left|\left(C_{\hat{\eta}_{ba}}^{k_{ab}},\Pi_{ba}\left(\{X_{a'b'}\}\right)C_{\xi_{ba}}^{k_{ab}}\right)\right|^2
\leq \left\|C_{\hat{\eta}_{ba}}^{k_{ab}}\right\|^2 \,\left\|\Pi_{ba}\left(\{X_{a'b'}\}\right)C_{\xi_{ba}}^{k_{ab}}\right\|^2
\leq 1
\eq
and thus $\Re S^{\Pi}_{ab}\leq 0$.
Since $\Re S^{EPRL}_{ab}\le 0$ as well,
$\Re \Sprop$ is maximal if $\Re S^{\Pi}_{ab}= 0$ and $\Re S^{EPRL}_{ab}= 0$.
According to {\asymptoticL}, $\Re S^{EPRL}_{ab} = 0$ implies
%
%
%
\begin{align}
\label{maxcrit}
\xi_{ab} = \frac{e^{i\phi_{ab}}}{\| Z_{ab}\|} X_a^{-1} z_{ab},
\text{ and } J\hat{\eta}_{ba} = \frac{e^{i\phi_{ba}}}{\|Z_{ba}\|} X_b^{-1} z_{ab}
\end{align}
for some set of phases $\phi_{ab}$ and $\phi_{ba}$.
Note that either of these determine $[z_{ab}] \in \CP^1$ uniquely.
They combine to yield
\begin{align}
\label{crit1}
X_a \xi_{ab}= \frac{\|Z_{ba}\|}{\|Z_{ab}\|} \ee{i\theta_{ab}} X_b J\hat{\eta}_{ba}
\end{align}
where $\theta_{ab}:= \phi_{ab} - \phi_{ba}$.
Let $\delta_{z_{ab}}$ be any variation of $z_{ab}$.
Since $S^\Pi$ is independent of $z_{ab}$ it follows that $\delta_{z_{ab}} \Sprop=\delta_{z_{ab}}  S^{EPRL}$,
which, from \asymptoticL, vanishes if and only if
\begin{align}
\label{crit2}
(X_a^\dagger)^{-1}\xi_{ab}= \frac{\|Z_{ab}\|}{\|Z_{ba}\|} \ee{i\theta_{ab}} (X_b^\dagger)^{-1} J\hat{\eta}_{ba} .
\end{align}
Inserting \eqref{crit1} into \eqref{crit2} and setting $\frac{\|Z_{ab}\|^2}{\|Z_{ba}\|^2}=\ee{r_{ba}}$,
one obtains the eigenvalue equation
\begin{align}
\label{eigen0}
X_{ba}^{\mathstrut} X_{ba}^\dagger J \eta_{ba} = e^{-r_{ba}} J\eta_{ba}^{\mathstrut}.
\end{align}
Applying $J$ to both sides then yields
\begin{align}
\label{eigen1}
X_{ba}^{\mathstrut}X_{ba}^{\dagger} \eta_{ba}^{\mathstrut}=\ee{r_{ba}} \eta_{ba}^{\mathstrut}.
\end{align}
These two equations uniquely determine $X_{ba}^{\mathstrut} X_{ba}^\dagger$.
Equation (\ref{eqn:ndef}) implies that 
$\xi$ is an eigenvector of 
$\mathbf{n}_\xi \cdot \pauli$ with eigenvalue $1$.
Using this, one checks that the following expression satisfies both
(\ref{eigen0}) and (\ref{eigen1}):
\begin{align*}
X_{ba}^{\mathstrut} X_{ba}^{\dagger}=\ee{2 i r_{ba} \boost_i\,\bm{n}_{\eta_{ba}}^i} .
\end{align*}
If $r_{ba}$ vanishes then $X_{ba}^{\mathstrut} X_{ba}^{\dagger}$ must be equal to the identity, which is excluded in the non-degenerate sector; otherwise
\begin{align}
\label{compon_n}
 \bm{n}_{\eta_{ba}}^i = P (\sgn \, r_{ba}) \tr\left(X_{ba}^{\mathstrut} {X}_{ba}^{\dagger}\pauli^i\right)
\end{align}
where $P$ is a positive real number.
\begin{lemma}
\label{lem:Pi_eigen}
If, for given boundary data $\{k_{ab},\xi_{ab}\}_{a\neq b}$, $\{X_a,z_{ab},\eta_{ba}\}$ is a solution to
${\rm\, Re}\, \Sprop = 0$ and $\delta_z \Sprop = 0$,
then
\begin{align}
\label{eqn:Ceigen}
\Pi_{ba}(\{X_{a'b'}\}) C_{\hat{\eta}_{ba}}^{k_{ab}} =
\Theta[(\sgn \, r_{ba})  \beta_{ab}(\{X_{a'b'}\}) k_{ab}] C_{\hat{\eta}_{ba}}^{k_{ab}}
\end{align}
where $\Theta[\cdot]$ is the Heaviside theta function.
%
%
%
\end{lemma}
{\startproof
Since $\Sprop$ is only defined in the non-degenerate sector, we have $X_{ba}^{\mathstrut} X_{ba}^\dagger \neq \mathrm{Id}$.
$\Re \Sprop = 0$ and $\delta_z \Sprop = 0$ then imply, by the above argument, that (\ref{compon_n}) holds.
But $C^{k_{ab}}_{\hat{\eta}_{ba}}$ is an eigenstate of
$\bm{n}_{\eta_{ba}}^i\rotoperator_i$ with eigenvalue $k_{ab}$,
so that (\ref{eqn:Ceigen}) is satisfied.
\finishproof}
In the same manner one can also derive an eigenvalue equation for $\xi_{ab}$ from equation \eqref{crit1} and \eqref{crit2}.
The resulting equations are the same as (\ref{eqn:Ceigen}) except that $a$ and $b$ are exchanged, and $\eta_{ba}$
is replaced by $\xi_{ab}$. One then directly reads off
\begin{align}
\label{sgnr}
\mathrm{sgn}\left[ \tr\left(X_{ab}^{\mathstrut} X_{ab}^{\dagger}\pauli^i\right)\;\bm{n}_{\xi_{ab}}^i\right]
=
\mathrm{sgn}\left[ \tr\left(X_{ba}^{\mathstrut} X_{ba}^{\dagger}\pauli^i\right)\,\bm{n}_{\eta_{ba}}^i\right]
= \sgn\,  r_{ba} ~.
\end{align}
This is relevant for the consistency of the following definition.
\begin{definition}
In the following we say that a given set of quantum data $\{\xi_{ab}, X_a, z_{ab}, \eta_{ba}\}_{a<b}$ is oriented if \eqref{crit1} and \eqref{crit2} holds. It is additionally proper oriented if
\begin{align}
\label{propor}
\beta_{ab}(\{\hat{X}_{a'}\})\bm{n}_{\xi_{ab}}^i \tr\left(X_{ab}^{\mathstrut} X_{ab}^{\dagger}\pauli_i\right)>0
\end{align}
for all $a,b$ with $a\neq b$
\end{definition}
\begin{lemma}
\label{lem:maximal}
Suppose for a given set of boundary data $\{k_{ab}, \xi_{ab}\}_{a\neq b}$
there exists integration variable values $\{X_a,[z_{ab}],[\eta_{ba}]\}_{a<b}$ such that
${\rm Re} \Sprop$ is maximal and $\delta_{z} \Sprop=0$. Then $\{\xi_{ab},X_a\}_{a\neq b}$
obeys orientation and proper orientation, and
\begin{align}
\label{etacrit}
[\eta_{ba}]=[\xi_{ba}].
\end{align}
\end{lemma}
{\startproof

As shown above, $\Re \Sprop = 0$ and $\delta_z \Sprop = 0$ imply (\ref{crit1}) and (\ref{crit2}),
which, by {\asymptoticL}, implies orientation.
Furthermore, by lemma \ref{lem:Pi_eigen}, (\ref{eqn:Ceigen}) holds.
Maximality then precludes the left hand side of (\ref{eqn:Ceigen}) from vanishing so that (\ref{eqn:Ceigen}) implies also proper orientation,
which in turn implies that $\Pi_{ba}$ acts as the identity on $C_{\hat{\eta}_{ba}}^{k_{ab}}$, whence
\begin{align*}
{\rm Re} S_{ab}^\Pi = \log \left| (C^{k_{ab}}_{\hat{\eta}_{ba}}, C^{k_{ab}}_{\xi_{ba}})\right|.
\end{align*}
But this in turn is maximal if and only if $C^{k_{ab}}_{\hat{\eta}_{ba}} \propto C^{k_{ab}}_{\xi_{ba}}$,
which holds if and only if $\eta_{ba}$ and $\xi_{ba}$ are equal up to rescaling by a complex number.

\finishproof}
\subsubsection{Stationarity with respect to \texorpdfstring{$X$}{X} and \texorpdfstring{$\eta$}{eta}}
\label{sec:Xeta}

We now examine the variation of the action with respect to the rest of the integration variables,
namely the group variables and the spinors $\eta$.
Let $\delta_X$ be an arbitrary variation of the group elements. Then
\bq
\begin{split}
\delta_X \exp(\Sprop)
&=\left(\delta_X \exp(S^{EPRL})\right) \exp(S^{\Pi})\\
&+ \exp(\Sprop)
 \sum_{a<b}\left(C_{\hat{\eta}_{ba}}^{k_{ab}}\,[\delta_X\,\Pi_{ba}\left(\{X_{a'b'}\}\right)]\,C_{\xi_{ba}}^{k_{ab}}\right)\,
\left[\left(C_{\hat{\eta}_{ba}}^{k_{ab}},\,\Pi_{ba}\left(\{X_{a'b'}\}\right)\,C_{\xi_{ba}}^{k_{ab}}\right)\right]^{-1}
~.
\end{split}
\eq
According to lemma \ref{lem:Pi_eigen} and lemma \ref{lem:maximal},
$C^{k_{ab}}_{\hat{\eta}_{ba}}$ and $C^{k_{ab}}_{\xi_{ba}}$ are eigenstates of $\Pi_{ba}$ with eigenvalue 1 if \eqref{crit1} and \eqref{crit2} are satisfied.  Corollary 11 in Appendix C of
{\jonProp} therefore implies
\begin{align}
\label{eqn:Xvar}
\left(C_{\hat{\eta}_{ba}}^{k_{ab}},\,[\delta_X\,\Pi_{ba}\left(\{X_{a'b'}\}\right)]\,C_{\xi_{ba}}^{k_{ab}}\right)=0
\end{align}
for all $a,b$. Thus, the variation with respect to the group elements just gives the original condition $\delta_X S^{EPRL}=0$
which is equivalent to closure of $(k_{ab}, \xi_{ab}, \hat{\eta}_{ba})$ {\asymptoticL},
and hence by (\ref{etacrit}) closure of $(k_{ab}, \xi_{ab}, \xi_{ba})$.
That is, variation with respect to the group elements again yields only
a condition on the boundary data.

It remains to consider variations $\delta \equiv \delta_{\eta_{ba}}$  with respect to each variable
$\eta_{ba}$, $a<b$:
\begin{align}
\label{A}
\begin{split}
\delta_{\eta_{ba}} \Sprop &=\delta_{\eta_{ba}}S_{ab}^{EPRL}+\delta_{\eta_{ba}} S_{ab}^\Pi\\[3pt]
&=2 k_{ab} \;\frac{\scal{J\delta\hat{\eta}_{ba},Z_{ba}}}{\scal{J\hat{\eta}_{ba},Z_{ba}}}
+
\frac{\delta_{\eta_{ba}}\left(C_{\hat{\eta}_{ba}}^{k_{ab}},\,\Pi_{ba}\left(\{X_{a'b'}\}\right)\,C_{\xi_{ba}}^{k_{ab}}\right)}{
\left(C_{\hat{\eta}_{ba}}^{k_{ab}},\,\Pi_{ba}\left(\{X_{a'b'}\}\right)\,C_{\xi_{ba}}^{k_{ab}}\right)}~.
\end{split}
\end{align}
At any point $p$ in the integration domain where $\Re \Sprop = 0$ and $\delta_z \Sprop = 0$,
we have the further simplification
\begin{align*}
\delta_{\eta_{ba}}\left(C_{\hat{\eta}_{ba}}^{k_{ab}},\,\Pi_{ba}\left(\{X_{a'b'}\}\right)\,C_{\xi_{ba}}^{k_{ab}}\right) |_p
&= \left(\delta_{\eta_{ba}} C_{\hat{\eta}_{ba}}^{k_{ab}}|_p,\,(\Pi_{ba}\left(\{X_{a'b'}\}\right)\,C_{\xi_{ba}}^{k_{ab}})|_p\right)\\
&= \left(\delta_{\eta_{ba}} C_{\hat{\eta}_{ba}}^{k_{ab}}|_p,\, C_{\xi_{ba}}^{k_{ab}}|_p\right)
= \delta_{\eta_{ba}}\left( C_{\hat{\eta}_{ba}}^{k_{ab}},\, C_{\xi_{ba}}^{k_{ab}}\right)|_p \\
&= \delta_{\eta_{ba}} \langle \hat{\eta}_{ba}, \xi_{ba}\rangle^{2k_{ab}}|_p
\end{align*}
where lemmas \ref{lem:Pi_eigen} and \ref{lem:maximal} have been used in the second line,
and equation (3.10) from \cite{ez2015} has been used in the last line. Furthermore, at maximal points of $\Re \Sprop$
equation (\ref{maxcrit}) uniquely determines $[z_{ab}]\in \CP^1$ in terms of $[\eta_{ba}]$.  Inserting these
identities into (\ref{A}) implies that,
when $\Re \Sprop = 0$ and $\delta_z \Sprop = 0$,
\begin{align}
\nonumber
\delta_{\eta_{ba}}\Sprop &=
2 k_{ab} \;\frac{\scal{J\delta\hat{\eta}_{ba},J\hat{\eta}_{ba}}}{\scal{J\hat{\eta}_{ba},J\hat{\eta}_{ba}}}
+
\frac{\delta_{\eta_{ba}}\scal{\hat{\eta}_{ba},\xi_{ba}}^{2k_{ab}}}{
\scal{\hat{\eta}_{ba},\xi_{ba}}^{2k_{ab}}}
=2 k_{ab} \;\frac{\scal{\hat{\eta}_{ba},\delta\,\hat{\eta}_{ba}}}{\scal{\hat{\eta}_{ba},\hat{\eta}_{ba}}}
+2 k_{ab}
\frac{\scal{\delta\,\hat{\eta}_{ba},\xi_{ba}}}{\scal{\hat{\eta}_{ba},\xi_{ba}}} \\
\nonumber
&=2 k_{ab} \;\frac{\scal{\hat{\eta}_{ba},\delta \, \hat{\eta}_{ba}}}{\scal{\hat{\eta}_{ba},\hat{\eta}_{ba}}}
+2 k_{ab} \frac{\scal{\delta\,\hat{\eta}_{ba},\hat{\eta}_{ba}}}{\scal{\hat{\eta}_{ba},\hat{\eta}_{ba}}}
= 2 k_{ab} \; \delta\scal{\hat{\eta}_{ba}, \hat{\eta}_{ba}} = 0,
\end{align}
whence the last critical point equation, $\delta_{\eta_{ba}} \Sprop = 0$, imposes no further conditions.
Let us summarize:
\begin{theorem}
Given boundary data $\{k_{ab},\xi_{ab}\}$,
$\{X_a,[z_{ab}], [\eta_{ba}]\}$ is a critical point of $\Sprop$ in the non-degenerate sector (i.e.  $X_{ab}\nin\SU(2)$ for all $a,b$) iff closure, orientation and proper orientation are satisfied,
and, for each $a<b$, $[\eta_{ba}] = [\xi_{ba}]$ and $[z_{ab}] \in \CP^1$ is as uniquely determined by
(\ref{maxcrit}).
\end{theorem}

\subsection{Asymptotics of the proper vertex}
\label{sec:main_result}

In order to apply the stationary phase method, the critical points need to be isolated.
There is only one continuous symmetry of the critical points, which is a  global $\SL$ symmetry.
This symmetry has been broken in the usual way by the insertion of $\delta(X_0)$ in the
vertex amplitude integral, so that all critical points are isolated.

In section \ref{sec:critical_points} it was proven that the critical points of the proper vertex in the non-degenerate sector are a subset of the critical points of the original EPRL model, specifically those which additionally satisfy proper orientation (\ref{propor}). In \cite{ez2015} it was shown that proper orientation holds if and only if the Plebanski 2-form reconstructed from the quantum data is in the Einstein-Hilbert sector.
However, in the analysis of Plebanski sectors performed in \cite{ez2015}
it was found that the only critical points of the original EPRL model
corresponding to the Einstein-Hilbert sector are those which give rise to the first term in (\ref{eqn:asymL}).
These critical points exist if and only if the boundary data is Regge-like and glues to a non-degenerate Lorentzian
4-simplex.
Furthermore, in this case there are no critical points in the degenerate sector,
so that the above analysis (and the analysis in the next section) applies.
Additionally, if the boundary data is not a vector geometry,
there are no critical points at all,
and hence in particular no critical points in the degenerate sector,
so that the above analysis again applies.
Finally, at the critical points of
$\Sprop$, by evaluating using the representative $\eta_{ba} = \xi_{ba}$ of $[\eta_{ba}] = [\xi_{ba}]$,
one sees that the action $\Sprop$ reduces to $S^{EPRL}$.
This yields the following:
\begin{theorem}[Proper EPRL-asymptotics]
\label{th:propasym}
Let $\{k_{ab},\bm{n}_{ab}\}$ be a set of non-degenerate, Regge-like boundary data
that glues to a Lorentzian 4-simplex  
and
$\psi^{\text{Regge}}_{\lambda k_{ab},\bm{n}_{ab}}$ the associated Regge state, then
\begin{align*}
\Aprop(\psi^{\rm Regge}_{\lambda k_{ab},\bm{n}_{ab}})\sim \left(\frac{1}{\lambda}\right)^{12} N^{\propdec}
\exp\left(i\lambda \gamma \sum_{a<b} k_{ab}\Theta_{ab}\right)
\end{align*}
with
\begin{align}
\label{Nprop}
N^{\propdec}
= \frac{(-1)^{\Xi} 2^{56} \pi^{12}}{\sqrt{\det H^{\propdec}|_{\rm crit}}}
\left(\frac{1+i\gamma}{1-i\gamma}\right)^5
\left(\prod_a \mu^{\rm Haar}_{X_a}|_{\rm crit} \right)
\prod_{a<b} \left(k_{ab}^2 \Omega_{ab}|_{\rm crit} \, \Omega_{\hat{\eta}_{ba}}|_{\rm crit} \right)
\end{align}
and where $\Theta_{ab}$ are the dihedral angles defined in section \ref{eprl_asympt}.
Here, $H^{\propdec}|_{crit}$ is the Hessian of $\Sprop$ and
$\mu_{X_a}^{\rm Haar}|_{\rm crit}$, $\Omega_{ab}|_{\rm crit}$, $\Omega_{\hat{\eta}_{ba}}|_{\rm crit}$
are the measure factors evaluated at any representative of the unique symmetry equivalence class of critical points,
with $\mu^{\rm Haar}_{X_a}$ the Haar measure on $X_a$.

%
%

If $\{k_{ab},\bm{n}_{ab}\}$ is not a vector geometry, then the amplitude decays
faster than any inverse power of $\lambda$ with any choice of phase.
\end{theorem}
Note, in particular, that in the asymptotics of the proper vertex,  the first term of (\ref{eqn:asymL})
corresponding to the Einstein-Hilbert sector is isolated.
As in {\asymptoticL}, the Hessian factor $H^{\propdec}|_{\rm crit}$, and measure factors $\Omega_{ab}|_{\rm crit}$,
$\Omega_{\hat{\eta}_{ba}}|_{\rm crit}$ depend on a choice of coordinates on $\CP^1$,
a dependence which cancels in (\ref{Nprop}).

Let us close this section with a few remarks on the coefficient $\left(\frac{1}{\lambda}\right)^{12} N^{\propdec}$.
In general the coefficient for the large $\lambda$ expansion of
\bq
f(\lambda)=\int_D \dif x\;a(x) \,\ee{\lambda S(x)}
\eq
over a $n$ dimensional manifold $D$ at a critical point $x_0$ is
\bq
a(x_0) \left(\frac{2 \pi}{\lambda}\right)^{n/2}\frac{1}{\sqrt{ \det H(x_0)}}~.
\eq
By the insertion of the resolution of identity we introduced 10 new $\CP^1$
variables and thus 20 new dimensions so that in total we now have $64$ dimensions,
so that $n$ in the above formula has changed as compared to the original EPRL vertex.
The introduction of the auxiliary variables also contributes further measure factors.
These two differences together yield an extra factor of
\bq
 \left(\frac{2 \pi}{\lambda}\right)^{10} \prod_{a<b}\frac{2 \lambda k_{ab}}{\pi} (\Omega_{\hat{\eta}_{ba}})|_{crit}
 = 2^{20}\prod_{a<b} k_{ab}\Omega_{\hat{\eta}_{ba}}|_{crit}
\eq
where the approximation $d_{\lambda k_{ab}}\approx 2\lambda k_{ab}$ was used.
Lastly, the determinant of the Hessian $H^\propdec$ is different from that in the EPRL case.
The Hessian $H^\propdec$ and its determinant are calculated explicitly in the
paper \cite{cev2015}, where, remarkably, it is found that the extra factor arising from the
change in Hessian \textit{exactly} cancels the above factors, so that the
coefficient in the asymptotics of the proper vertex is the same as the coefficient
in front of the Einstein-Hilbert term in the EPRL case.

\section{Technical completion of the argument for the asymptotics}
\label{sec:completion}

In this section we give a more mathematically careful
argument for the asymptotics of the proper vertex derived in the previous section,
arriving at the same conclusion.
We begin by deriving the asymptotic form of the integrand of the proper vertex, showing that
the projector part $S^\Pi$ of the action is asymptotic, in the large spin limit,
to an action which scales linearly with the spins.
The general strategy will then be to express the asymptotic limit of the vertex amplitude
as the integral of this asymptotic integrand, to which the extended stationary phase theorem and other
related arguments can be applied.

\subsection{Asymptotic expression for projector part of the action}
\label{sec:projasym}

The action $\Sprop$ for the proper vertex is given by (\ref{eqn:propaction}):
\begin{align}
\nonumber
\Sprop[\lambda; \{k_{ab},\xi_{ab}\}; \{X_a, z_{ab}, \eta_{ba}\}]
& := \Sprop[\{\lambda k_{ab},\xi_{ab}\}; \{X_a, z_{ab}, \eta_{ba}\}] \\
\label{stph_proper}
& = \lambda S^{EPRL}[\{k_{ab}, \xi_{ab}, \hat{\eta}_{ba}\}; \{X_a, z_{ab}\}]
+ \sum\limits_{a<b} S^\Pi_{ab}
\end{align}
where
\begin{align*}
S^\Pi_{ab}[\{\lambda k_{ab}, \xi_{ba}\}; \{X_a,\eta_{ba}\}]
= \mathrm{log}\left(C_{\hat{\eta}_{ba}}^{k_{ab}},
\Pi_{ba}\left(\{X_{a'b'}\}\right)\,C_{\xi_{ba}}^{k_{ab}}\right)~.
\end{align*}
We begin by deriving the asymptotic expression for $S^{\Pi}_{ab}$.
For each $a<b$, let $\nu_{ba}$ be any normalized spinor such that
$\mathbf{n}_{\nu_{ba}} = \beta_{ab}(\{X_{a'}\}) \frac{\tr(X_{ba}^{\mathstrut} X_{ba}^\dagger \pauli)}{|\tr(X_{ba}^{\mathstrut} X_{ba}^\dagger \pauli)|}$
(see (\ref{eqn:ndef})).
For any normalized spinor $\xi$, let
\begin{align*}
g(\xi) := \left(\begin{array}{cc} \xi_0 & -\overline{\xi}_1\\ \xi_1 & \overline{\xi}_0 \end{array}\right) \in \SU(2) .
\end{align*}
Then
\begin{align*}
\ket{\xi;k,m} := g(\hat{\xi}) \ket{k,m}
\end{align*}
is an eigenstate of $\mathbf{n}_\xi \cdot \rotoperator$ with eigenvalue $m$ in the spin $k$ representation,
and the coherent states $C^k_\xi$ introduced earlier are simply $\ket{\xi; k,k}$.
One then has the following explicit expression for the projector:
\begin{align*}
\Pi_{ba}(\{X_{a'}\}) = \Pi_{(0,\infty)}(\mathbf{n}_{\nu_{ba}[X]} \cdot \rotoperator)
= \sum_{m>0}^{k_{ab}} \ket{\nu_{ba}[X]; k_{ba}, m} \bra{\nu_{ba}[X]; k_{ba}, m}
\end{align*}
so that
\begin{align*}
e^{S^{\Pi}_{ab}} &:=
\bra{\eta_{ba};k_{ab}, k_{ab}} \Pi_{ba}(\{X_{a'}\}) \ket{\xi_{ba}; k_{ab}, k_{ab}}\\
&= \sum_{m>0}^{k_{ab}} \langle \eta_{ba}; k_{ab}, k_{ab} | \nu_{ba}[X]; k_{ab}, m \rangle
\langle \nu_{ba}[X]; k_{ab}, m | \xi_{ba}; k_{ab}, k_{ab} \rangle \\
&= \sum_{m>0}^{k_{ab}} \langle k_{ab}, k_{ab}| g(\hat{\eta}_{ba})^{-1} g(\nu_{ba}[X]) | k_{ab}, m \rangle
\langle k_{ab}, m | g(\nu_{ba}[X])^{-1} g(\xi_{ba}) |k_{ab}, k_{ab} \rangle.
\end{align*}
Given unit spinors $\eta$ and $\nu$,
\begin{align*}
g(\eta)^{-1} g(\nu) = \left(\begin{array}{cc}\langle \eta, \nu\rangle & \langle \eta, J\nu \rangle \\
\langle J \eta, \nu \rangle & \langle \nu, \eta \rangle \end{array} \right).
\end{align*}
This, together with equation (32.1.5) in \cite{ThiemannBook}
yields
\begin{align*}
e^{S^{\Pi}_{ab}} &= \sum^{k_{ab}}_{m>0} \begin{pmatrix} 2k_{ab} \\ k_{ab}+m \end{pmatrix} (\scal{\hat{\eta}_{ba},\nu_{ba}[X]} \scal{\nu_{ba}[X],\xi_{ba}})^{k_{ab}+m} (\scal{\hat{\eta}_{ba},J\nu_{ba}[X]} \scal{J\nu_{ba}[X],\xi_{ba}})^{k_{ab}-m}~.
\end{align*}
Let $x_{ab} := \scal{\hat{\eta}_{ba},\nu_{ba}} \scal{\nu_{ba},\xi_{ba}}$ and $y_{ab} := \scal{\hat{\eta}_{ba},J\nu_{ba}} \scal{J\nu_{ba},\xi_{ba}}$.
In the following we temporarily suppress indices $(ab)$.
We then have
\begin{align}
\label{eqn:Ainxy}
A := e^{S^\Pi_{ab}} = \sum^{k}_{m>0} \binom{2k}{k+m} x^{k+m} y^{k-m} = \sum^{k}_{m>0} \begin{pmatrix} 2k \\ k-m \end{pmatrix} y^{k-m} x^{k+m} .
\end{align}
At this point, in order to proceed, we first consider the case in which $k$ is an integer.
Setting $i=k-m$, the above equation becomes
\begin{align}
\label{eqn:tayrem}
A = \sum^{k-1}_{i=0} \begin{pmatrix} 2k \\ i \end{pmatrix} y^{i} x^{2k-i}
= (x+y)^{2k} - \sum^k_{i=0} \begin{pmatrix} 2k \\ i \end{pmatrix} y^{2k-i} x^i .
\end{align}
This is a remainder in a $k$-th order Taylor expansion of the function $(y+tx)^{2k}$ around $t=0$ and evaluated at $t=1$.
Using the integral formula for the remainder, we have
\begin{align*}
A = \int_0^1 \frac{1}{k!} \frac{(2k)!}{(k+1)!}(y+tx)^{k-1} x^{k+1} (1-t)^k dt
= \begin{pmatrix} 2k \\ k \end{pmatrix} x^{k+1} \int^1_0 (y+(1-t)x)^{k-1} kt^k dt .
\end{align*}
Then changing variables $s=t^{k+1}$ and introducing $z=x/y$ we get:
\begin{align}
\label{eqn:tayint}
A =  \begin{pmatrix} 2k \\ k \end{pmatrix}  \frac{k}{k+1} (xy)^k z \int^1_0 \left(1+\left(1-s^{\frac{1}{k+1}}\right)z\right)^{k-1} ds
\end{align}
Define
\begin{align*}
f_k(s,z) = \left(1+\left(1-s^{\frac{1}{k+1}}\right)z\right)^{k-1} .
\end{align*}
To study asymptotics of the integral we will apply the dominated convergence theorem to the sequence of functions $\{f_k\}$. This sequence converges pointwise to the function $s^{-z}$. Furthermore, for $\abs{z}<1$ we have
\begin{align*}
\abs{f_k(s,z)} < \left(1+\left(1-s^{\frac{1}{k+1}}\right)\abs{z}\right)^{k-1}
\end{align*}
Now we make use of the following two inequalities which can be derived from $e^x\geq1+x$:
\begin{align*}
1-s^\frac{1}{k+1} \leq -\frac{1}{k+1} \log{s}, \qquad
\left(1+ \frac{a}{k+1}\right)^{k-1} \leq (e^{\frac{a}{k+1}})^{k-1} \leq e^a \text{ for }a>0
\end{align*}
these give
\begin{align*}
\abs{f_k(s,z)} < \left(1-\frac{\abs{z}\log{s}}{k+1}\right)^{k-1} \leq \exp(-\abs{z}\log{s}) = s^{-\abs{z}} .
\end{align*}
Thus we obtain an integrable bound for the function sequence $\{f_k\}$ when $\abs{z}<1$.
Using the dominated convergence theorem and applying Stirling's formula we then get from \eqref{eqn:tayint}:
\begin{align*}
A \sim \frac{(4xy)^k}{\sqrt{\pi k}} \frac{k}{k+1} \frac{z}{1-z}
\sim \frac{(4xy)^k}{\sqrt{\pi k}} \frac{x}{y-x}
\end{align*}
If $\abs{z}>1$, we can write the second term in \eqref{eqn:tayrem} as a Taylor remainder:
\begin{align*}
A = (x+y)^{2k} - \begin{pmatrix} 2k \\ k \end{pmatrix} (xy)^k \int^1_0 \left(1+\left(1-s^{\frac{1}{k}}\right)\frac{1}{z}\right)^k ds
\end{align*}
In this case we define $w=1/z$ and consider the sequence of functions $\{g_k\}$ with $\abs{w}<1$:
\begin{align*}
g_k(s,w) = \left(1+\left(1-s^{\frac{1}{k}}\right)w\right)^k
\end{align*}
This function sequence converges pointwise to $s^{-w}$. We can also obtain an integrable bound $s^{-\abs{w}}$ when $\abs{w}<1$ by an argument essentially similar to above.
Hence, we can again use the dominated convergence theorem and we get the following asymptotic expression:
\begin{align*}
A \sim (x+y)^{2k} - \frac{(4xy)^k}{\sqrt{\pi k}} \frac{1}{1-w}
\sim (x+y)^{2k} + \frac{(4xy)^k}{\sqrt{\pi k}} \frac{x}{y-x}
\end{align*}
Thus if $\abs{x+y}^2\geq\abs{4xy}$ we obtain $A \sim (x+y)^{2k}$, and if $\abs{x+y}^2<\abs{4xy}$ we get
\begin{align*}
A \sim  \frac{(4xy)^k}{\sqrt{\pi k}} \frac{x}{y-x}
\end{align*}
If we restore the subscripts and summarize all the cases, we obtain
\begin{align}
\label{eqn:intasym}
\exp (S^\Pi_{ab}) \sim
\left\{ \begin{array}{ll}
(x_{ab}+y_{ab})^{2 \lambda k_{ab}} & \text{ if }|x_{ab}|>|y_{ab}|\text{ and }|x_{ab}+y_{ab}|^2\ge|4x_{ab}y_{ab}| \\
\frac{(4 x_{ab} y_{ab})^{\lambda k_{ab}}}{\sqrt{\pi \lambda k_{ab}}}\frac{x_{ab}}{y_{ab}-x_{ab}} & \text{ if }|x_{ab}|<|y_{ab}|\text{ or }|x_{ab}+y_{ab}|^2<|4x_{ab}y_{ab}|
\end{array} \right.
\end{align}
as long as $\{X_a\}$ is not in the degenerate sector,
and $|x_{ab}| \neq |y_{ab}|$.

Up until now we have assumed that $k_{ab}$ is an integer. In the case when $k_{ab}$ is restricted to be half integer, through exactly the same
sort of argument, we have the following, almost identical asymptotics:
\begin{align}
\label{eqn:halfintasym}
\exp (S^\Pi_{ab}) \sim
\left\{ \begin{array}{ll}
(x_{ab}+y_{ab})^{2 \lambda k_{ab}} & \text{ if }|x_{ab}|>|y_{ab}|\text{ and }|x_{ab}+y_{ab}|^2\ge|4x_{ab}y_{ab}| \\
\frac{(4 x_{ab} y_{ab})^{\lambda k_{ab}}}{\sqrt{\pi \lambda k_{ab}}}
\frac{(x_{ab}y_{ab})^{1/2}}{y_{ab}-x_{ab}} &
\text{ if }|x_{ab}|<|y_{ab}|\text{ or }|x_{ab}+y_{ab}|^2<|4x_{ab}y_{ab}|
\end{array} \right.
\end{align}
The only difference lies in the exponent of a single factor of $x_{ab}$.
Nevertheless, as a consequence of this difference, in order to have a well-defined asymptotic limit of the
integrand, the asymptotic limits in which certain spins $k_{ab}$ are integer or half-integer must be
considered separately --- there are $2^{10}$ such different asymptotic limits to consider,
according to whether each of the ten spins $k_{ab}$ is integer or half-integer.
However, it will turn out that in all of these cases, \textit{exactly the same arguments
can be applied, yielding exactly the same final asymptotics for the vertex amplitude integral}.
For this reason, it suffices to restrict consideration to the case in which all of the spins $k_{ab}$
are integer, and we do so for the rest of this paper.  At the end of the subsection we will make
one final remark on the other cases.

For the case in which all spins $k_{ab}$ are integer, the above result (\ref{eqn:intasym})
leads us to define the following functions of the integration variables.
\begin{itemize}
\item
$\tilde{S}(X_a, z_{ab}, \eta_{ba}):=
S^{EPRL}[\{k_{ab}, \xi_{ab}, \hat{\eta}_{ba}\}; \{X_a, z_{ab}\}]
+ \sum\limits_{a<b} \tilde{S}^\Pi_{ab}$,
where
\begin{align}
\label{eqn:SPabt}
\tilde{S}^\Pi_{ab}:= \left\{
\begin{array}{ll}
2 k_{ab} \log (x_{ab}+y_{ab}) & \text{ if }|x_{ab}|>|y_{ab}|\text{ and }|x_{ab}+y_{ab}|^2\ge|4x_{ab}y_{ab}| \\
k_{ab} \log (4 x_{ab} y_{ab}) & \text{ if }|x_{ab}|\le|y_{ab}|\text{ or }|x_{ab}+y_{ab}|^2<|4x_{ab}y_{ab}|
\end{array}
\right.
\end{align}

\item
$\tilde{B}(\lambda):= \prod_{a<b} \tilde{B}_{ab}(\lambda)$
where
\begin{align}
\label{eqn:Babt}
\tilde{B}_{ab}(\lambda) := \left\{
\begin{array}{ll}
1 & \text{ if }|x_{ab}|>|y_{ab}|\text{ and }|x_{ab}+y_{ab}|^2\ge|4x_{ab}y_{ab}| \\
(\pi \lambda k_{ab})^{-1/2} & \text{ if }|x_{ab}|\le|y_{ab}|\text{ or }|x_{ab}+y_{ab}|^2<|4x_{ab}y_{ab}|
\end{array}
\right.
\end{align}

\item
$\tilde{\mu}(X_{a}, z_{ab}, \eta_{ba}):= \prod_{a<b} \tilde{\mu}_{ab}(X_{a'}, z_{a'b'}, \eta_{b'a'})$
where
\begin{align}
\label{eqn:muabt}
\tilde{\mu}_{ab}(X_{a'}, z_{a'b'}, \eta_{b'a'}) := \left\{
\begin{array}{ll}
1 & \text{ if }|x_{ab}|>|y_{ab}|\text{ and }|x_{ab}+y_{ab}|^2\ge|4x_{ab}y_{ab}| \\
\frac{x_{ab}}{x_{ab} - y_{ab}} & \text{ if }|x_{ab}|\le|y_{ab}|\text{ or }|x_{ab}+y_{ab}|^2<|4x_{ab}y_{ab}|
\end{array}
\right.
\end{align}
\end{itemize}
Define
\begin{itemize}
\item $M_X := \SL^4 = \{(X_a)\}$, $M_{\eta z}:= (\CP^1)^{20} = \{(\eta_{ba}, z_{ba})\}$
\item $M:= M_X \times M_{\eta z}$
\item $\mathcal{D}:= \{ p \in M_X | X_{ab}(p) \in \SU(2)\text{ for some }a<b\}$, the degenerate sector.
\item $\tilde{M}:= (M_X \setminus \mathcal{D})\times M_{\eta z}$
\item $\mathcal{E}:= \{ p \in \tilde{M} | \,\,\,\, |4x_{ab}y_{ab}| = |x_{ab} + y_{ab}|^2 \text{ or }
|x_{ab}| = |y_{ab}| \text{ for some }a<b\}$
\item $\mathcal{F}:= \{ p \in \tilde{M} | \Re \tilde{S} \text{ diverges to }
-\infty\text{ as p is approached}\}$\footnote{Singular
points of this type also appear in the original EPRL action \asymptoticL; the authors of {\asymptoticL}
did not address these points
because they trivially have no effect on the asymptotics.
We address them due to the need for more detailed analysis in our case.
For clarity, note that, in our case, even though by definition
$\mathcal{F}$ contains no points in the degenerate sector, it does seem to contain points
in every neighborhood of every point of the degenerate sector;  this prompted us to be especially careful
about handling these sets in the argument that follows.
}.
\item $\mathfrak{S}:= \left(\mathcal{D} \times M_{\eta z}\right)
\cup \mathcal{E} \cup \mathcal{F}$.

\end{itemize}
We then have
\begin{align}
\label{eqn:asymS}
\exp(\Sprop(\lambda k_{ab}, \xi_{ab}, \xi_{ba}; X_a, z_{ab}, \eta_{ba}))
\sim \tilde{B}(\lambda) \tilde{\mu} \exp(\lambda \tilde{S})
\end{align}
throughout $\tilde{M}$, except in the set of measure zero $\mathfrak{S}$.

In the above definitions (\ref{eqn:SPabt}-\ref{eqn:muabt}), we have extended the quantities
$\tilde{S}$, $\tilde{B}$, $\tilde{\mu}$
to the case where $|x_{ab}|=|y_{ab}|$ for some $a<b$.
This extension is smooth everywhere except where $|4x_{ab}y_{ab}|=|x_{ab}+y_{ab}|^2$
for some $a<b$. 
%
%
%
More importantly, the real part of $\tilde{S}$ is, by construction, continuous
even at points where $|x_{ab}+y_{ab}|^2 = |4x_{ab}y_{ab}|$ for some $a<b$,
so that $\Re \tilde{S}$ is continuous throughout $\tilde{M}$.
This latter fact will be used in subsection \ref{sec:secondterm}.

Let us close with a final remark on the case in which one or more of the spins $k_{ab}$ are half integer.
From (\ref{eqn:halfintasym}), the only definition above that changes in this case
is the definition of the measure factor $\tilde{\mu}_{ab}$, which becomes
\begin{align*}
\tilde{\mu}_{ab} = \left\{
\begin{array}{ll}
1 & \text{ if }|x_{ab}|>|y_{ab}|\text{ and }|x_{ab}+y_{ab}|^2\ge|4x_{ab}y_{ab}| \\
\frac{(x_{ab}y_{ab})^{1/2}}{x_{ab} - y_{ab}} & \text{ if }|x_{ab}|\le|y_{ab}|\text{ or }|x_{ab}+y_{ab}|^2<|4x_{ab}y_{ab}|
\end{array}
\right. .
\end{align*}
As already noted, if the measure factor (\ref{eqn:muabt}) is replaced with the above measure factor,
every element of the argument which follows in the rest of this paper will continue to apply, and,
in fact, the same answer will be found for the asymptotics of the vertex amplitude.
More specifically, the measure factor above shares all of the same properties regarding smoothness
which will be needed for the arguments which
follow, and both measure factors $\tilde{\mu}_{ab}$ evaluate to the same value at critical points of $\Sprop$, namely $1$.
As a consequence it suffices to consider only the case in which all spins $k_{ab}$ are restricted to be integer.
We do so from now on.

\subsection{Set up and separation into two terms}

Let $\mu[X_a, \eta_{ba}, z_{ab}]$ denote the measure factor in equation (\ref{eqn:newExP}),
so that the proper vertex amplitude takes the form
\begin{align*}
\Aprop(\lambda) = \int_M \mu e^{\Sprop}.
\end{align*}
The stationary phase theorems in \cite{hormander1983}
--- which we will apply later to this integral with $e^{\Sprop}$ replaced by its asymptotic limit ---
technically require the measure factor in the integral
to have compact support.
Here, $\mu$ does not have compact support, due to the non-compactness of $\SL$.
%
%
This is also true in the original analysis of the Lorentzian EPRL asymptotics {\asymptoticL}:
The relevant integral is again a non-compact integral over copies of $\SL$,
yet the theorems of \cite{hormander1983} are applied
%
%
treating it more or less as a technical detail.
Indeed, in physics, stationary phase methods are routinely applied to non-compact
integrals where the non-compactness is due to an ``infinite'' direction in the integration domain,
as is the case here.
We assume we can do the same in the present context.
Furthermore, it will be useful to formulate this assumption explicitly due to the additional
subtleties of the present case.

To be precise,
we introduce a sequence of bump functions $u_I$, $I \in \mathbb{N}$, such that
\begin{enumerate}
\item Each $u_I$ equals $1$ in a neighborhood of each critical point of $\Sprop$
\item Each $u_I$ is smooth and has compact support in $M$
\item For each $p \in M$, $u_I(p)$ is non-decreasing
%
%
with $I$, and there exists some
$N_p$ such that $u_I(p)=1$ for all $I>N_p$.
\end{enumerate}
Such a sequence of bump functions has nested supports, with each $u_I$
equal to $1$ in a region larger than that in which $u_{I-1}$ is equal to $1$, until,
as $I \rightarrow \infty$, $u_I = 1$ on the entire manifold $M$.
%
%
%
The proper vertex can then be expressed as
\begin{align*}
\Aprop(\lambda) = \lim_{I \rightarrow \infty} \Apropi{I}(\lambda),
\qquad \text{where} \qquad
\Apropi{I}(\lambda) := \int_M u_I \mu e^{\Sprop}.
\end{align*}
In the rest of this subsection and the next, we study the asymptotics of the integral
$\Apropi{I}(\lambda)$ for a given $I$.
We will prove that, \textit{for every} $I$, $\Apropi{I}(\lambda)$
\textit{has the asymptotics derived in theorem \ref{th:propasym}.} The key assumption is then the following:
\textit{We assume that, in the case of $\Apropi{I}$, the asymptotic limit $\lambda \rightarrow \infty$
commutes with the limit $I \rightarrow \infty$.}
%
%
%
This is our precise formulation of the assumption that the non-compactness of
$\SL$ does not interfere with the application of the stationary phase
methods.
%
%
%
From this assumption it will then follow that
\textit{The asymptotics of $\Aprop$ are those derived in theorem \ref{th:propasym}.}

For the rest of this subsection and the next, let $I$ be arbitrary and fixed.
We have the following lemma.
\begin{lemma}
The set of critical points of $\Sprop$ is separated from
$\mathfrak{S} \equiv \left(\mathcal{D} \times M_{\eta z}\right)
 \cup \mathcal{F} \cup \mathcal{E}$.
\end{lemma}
{\startproof
Denote the set of critical points of $\Sprop$ by $\Crit(\Sprop)$.
That there are no critical points of $\Sprop$ in the degenerate sector or in $\mathcal{F}$
is immediate, as $\Sprop$ is not even defined there.
To see that there are no critical points in $\mathcal{E}$,
note that the critical point equations of $\Sprop$ derived in section \ref{sec:propasym}
imply that $|x_{ab}|=1$ and $y_{ab}=0$ for all $a<b$, so that neither
$|4x_{ab}y_{ab}|=|x_{ab}+y_{ab}|^2$ nor $|x_{ab}|=|y_{ab}|$ is satisfied for any $a<b$, so that
$\mathcal{E}$ is excluded.
Furthermore, since $\mathfrak{S}$ and $\Crit(\Sprop)$ are closed sets in $\tilde{M}$,
not only is $\Crit(\Sprop)$
disjoint from $\mathfrak{S}$, it is separated from $\mathfrak{S}$ in $\tilde{M}$.
\finishproof}
This fact makes it possible to separate the integral into two terms:
One which includes all critical points of $\Sprop$
and can be shown to yield the asymptotics given in section \ref{sec:propasym}, and another which covers $\mathfrak{S}$
and which can be shown to be sub-dominant.

We accomplish the separation into two terms
using a partition of unity which we construct with some explicitness for technical reasons relevant in
subsection \ref{sec:secondterm}. For this purpose, we introduce the following.
\begin{quote}
{\it
Let  $\varphi:\R \rightarrow \R$ be an arbitrary but fixed smooth function such that
\begin{enumerate}
\item $\varphi(x) = 1$ for $x>3/4$, and
\item $\varphi(x) = 0$ for $x<1/4$.
\end{enumerate}
From these properties and smoothness follows furthermore that all order derivatives
of $\varphi(x)$ are uniformly bounded.
}
\end{quote}
Because they are disjoint and closed in $M$,
$\Crit(\Sprop)$ and $\mathcal{D} \times M_{\eta z}$
are separated in $M$.
It follows that there
exists a smooth function $\varphi_\mathcal{D}:\tilde{M} \rightarrow [0,1]$ such that
$\varphi_\mathcal{D} = 1$ in an open neighborhood of $\Crit(\Sprop)$ and $\varphi_\mathcal{D} = 0$ in an
open neighborhood of $\mathcal{D}$.
We then define $\rho_1, \rho_2: M \rightarrow [0,1]$ by
\begin{align*}
\rho_1 &:= \varphi_\mathcal{D} \cdot \varphi(\Re \tilde{S} + 1)
\cdot \prod_{a<b}\varphi\left(2-8|y_{ab}|\right)\varphi\left(|x_{ab}|-|y_{ab}|\right)\varphi\left( |x_{ab}+y_{ab}|^4 \!\!- \!|4x_{ab}y_{ab}|^2\right)\\
\rho_2 &:= (1 - \rho_1)
\end{align*}
These two functions have the following properties:
\begin{enumerate}

\item They form a partition of unity over $M$.

\item They are smooth throughout $M$.  To see this, recall that $\Re \tilde{S}$ is smooth everywhere except
where it diverges to minus infinity or where $|x_{ab}+y_{ab}|^2 = |4x_{ab}y_{ab}|$,
and $x_{ab}$ and $y_{ab}$ are smooth everywhere except in the degenerate sector.
In a neighborhood of any point where any of these are not smooth,
$\rho_1 = 0$, so that $\rho_1$ is smooth throughout $M$.
Hence $\rho_2$ is also smooth throughout $M$.

\item $\rho_1=1$ in an open neighborhood of each critical point of $\Sprop$,
%
%
the support of $\rho_1$ excludes an open neighborhood of $\mathfrak{S}$, and, within the support of $\rho_1$,
$|y_{ab}| \le 7/32< 1/4$.\footnote{so that $|y_{ab}|$ is uniformly bounded below and away from $1/4$, 
which will be used later on.}

\item Within the support of $\rho_1$, $|x_{ab}|>|y_{ab}|$ and $|x_{ab}+y_{ab}|^2>|4x_{ab}y_{ab}|$, so that $\tilde{B}(\lambda) \equiv 1$,
$\tilde{\mu} \equiv 1$, and $\tilde{S}^\Pi_{ab} \equiv 2 k_{ab} \log (x_{ab}+y_{ab})$ throughout this region.
%
%

\item \label{rhotwosupp} The support of $\rho_2$ excludes an open neighborhood of each critical point of $\Sprop$.
%
%
%

\end{enumerate}
We then have
\begin{align}
\label{eqn:twoterms}
\Apropi{I}(\lambda) = \int \rho_1 u_I \mu e^{\Sprop(\lambda)} + \int \rho_2 u_I \mu e^{\Sprop(\lambda)}.
\end{align}
In section \ref{sec:first}, we will prove that, for the first term in this expression, the asymptotic limit
commutes with the integral, allowing the standard stationary phase theorem \cite{hormander1983} to be used.
\textit{We will see that this yields the asymptotics given in the previous section of the paper,
section \ref{sec:propasym}.}
In section \ref{sec:secondterm}, we make the assumption that, for the second term, the asymptotic limit again commutes with the integral,
%
%
and show that this term falls off faster than the first term. For both of these sections, it will be useful to introduce the notation $o(f(\lambda))$, which denotes some function $g(\lambda)$ such that $\lim_{\lambda \rightarrow \infty}g(\lambda)/f(\lambda)=0$. 
%
%

\subsection{Equivalence of critical points of \texorpdfstring{$\Sprop$}{S(+)} and \texorpdfstring{$\tilde{S}$}{asymptotic action}}

For analyzing each of the two terms in (\ref{eqn:twoterms}),
it will be important that the critical points
of $\Sprop$ and $\tilde{S}$ are the same.  This is established by the following lemma.

\begin{lemma}
\label{lem:criteq}
The critical points of $\Sprop$ are the same as the critical points of $\tilde{S}$.
\end{lemma}
{\startproof

We begin by noting that, since $S^\Pi$ and $\tilde{S}^\Pi$ are both independent of $z$,
for any variation $\delta_z$ with respect to the $z_{ab}$ variables,
one has
\begin{align}
\label{eqn:zSasym}
\delta_z \Sprop = \delta_z \tilde{S}.
\end{align}
With this we proceed to prove that every critical point of $\tilde{S}$ is a
critical point of $\Sprop$ and vice versa.

\textit{Suppose $p=(X_a, z_{ab}, \eta_{ba})$ is a critical point of $\tilde{S}$.}
It follows that $0= \Re \tilde{S} = \Re S^{\rm EPRL} + \Re S^\Pi$. 
Since both terms on the right hand side are non-positive, it follows that
$\Re S^{EPRL} = 0$.
Additionally, (\ref{eqn:zSasym}) implies $\delta_z S^{EPRL} = 0$
for any variation $\delta_z$.
As shown in section \ref{sec:zcrit},
these two conditions imply that
equation (\ref{compon_n}) holds, so that $\mathbf{n}_{\eta_{ba}} = \pm \mathbf{n}_{ba}[X]$
whence $[\eta_{ba}] = [\nu_{ba}[X]]$ or $[\eta_{ba}] = [J \nu_{ba}]$, so that either $y_{ab} = 0$ or $x_{ab} = 0$.
From equation (\ref{eqn:SPabt}), as $x_{ab} \rightarrow 0$, $\Re \tilde{S} \rightarrow -\infty$,
which cannot be the case at $p$, as we have assumed $p$ a critical point of $\tilde{S}$.
Thus $y_{ab}=0$ for each $a<b$.  From (\ref{eqn:SPabt}) this implies $\tilde{S}^\Pi_{ab} = 2k_{ab} \log (x_{ab})$
whence the maximality of $\tilde{S}^\Pi_{ab}$ implies $x_{ab} = 1$, which implies $[\eta_{ba}]=[\xi_{ba}]$
for all $a<b$.  But then all of the critical point equations of $\Sprop$ derived in section \ref{sec:propasym} are satisfied,
so that $p$ is also a critical point of $\Sprop$.

\textit{Suppose conversely that $p=(X_a, z_{ab}, \eta_{ba})$ is a critical point of $\Sprop$.}
Then, from section \ref{sec:Xeta},
\begin{align}
\label{eqn:critcons}
[\xi_{ba}] = [\eta_{ba}]= [\nu_{ba}[X]], \qquad \text{so that}
\qquad x_{ab}=1\text{ and }y_{ab}=0 \qquad \text{for all }a<b.
\end{align}
Thus, in a neighborhood of $p$, for each $a<b$,
\begin{align}
\label{eqn:SPiform}
\tilde{S}^\Pi_{ab} = 2k_{ab} \log(x_{ab}+y_{ab}) = 2k_{ab} \log(\langle \hat{\eta}_{ba}, \xi_{ba} \rangle).
\end{align}
This allows one to prove the following equations:
\begin{align}
\label{eqn:ReSab}
\Re \tilde{S}^\Pi_{ab} & \eqat{p}
\Re S^\Pi_{ab} \eqat{p} 0\\
\label{eqn:dXSab}
\delta_X \tilde{S}^\Pi_{ab} &\eqat{p} \delta_X S^\Pi_{ab} \eqat{p} 0\\
\label{eqn:detaSab}
\delta_{\eta} \tilde{S}^\Pi_{ab} &\eqat{p} \delta_{\eta} S^\Pi_{ab}
\end{align}
where $\eqat{p}$ denotes equality at $p$,
and where $\delta_X$ and $\delta_{\eta}$
are any variations with respect to the $X_a$ and $\eta_{ba}$ variables, respectively.
The first follows by plugging equations (\ref{eqn:critcons}) into both sides.
The second follows because $\tilde{S}^\Pi$ in (\ref{eqn:SPiform})
is independent of $X$, together with equation (\ref{eqn:Xvar}).
The third follows from the calculation of $\delta_\eta S^\Pi_{ab}$ carried out in
section \ref{sec:Xeta},
together with explicit calculation of $\delta_\eta \tilde{S}^\Pi_{ab}$ from
(\ref{eqn:SPiform}), and evaluating on (\ref{eqn:critcons}).
Lastly, from (\ref{eqn:zSasym}), $\delta_z S^\Pi \eqat{p} \delta_z \tilde{S}^\Pi$.
These four equations imply
\begin{align*}
\Re \tilde{S} \eqat{p} \Re S \eqat{p} 0,
\qquad \delta_z \tilde{S} \eqat{p} \delta_z \Sprop \eqat{p} 0,
\qquad \delta_X \tilde{S} \eqat{p} \delta_X \Sprop \eqat{p} 0,
\qquad \delta_{\eta} \tilde{S} \eqat{p} \delta_{\eta} \Sprop \eqat{p} 0,
\end{align*}
so that $p$ is also a critical point of $\tilde{S}$.
\finishproof}

\subsection{Evaluation of the first term}
\label{sec:first}


\begin{lemma}
\label{lem:diffprod}
Given any two sequences of complex numbers $a_i$ and $\tilde{a}_i$, $i=1, \dots n$, we have
\begin{align}
\label{eqn:diffprod}
\prod_{i=1}^n a_i - \prod_{i=1}^n \tilde{a}_i
= \sum_{i=1}^n \left(\prod_{j=1}^{i-1} a_j\right)\left(\prod_{k=i+1}^n \tilde{a}_k \right)
(a_i - \tilde{a}_i).
\end{align}
where, when the upper limit of the product symbol is less than the lower limit, the product
is understood to be $1$.
\end{lemma}
{\startproof

We proceed by induction in $n$.
It is immediate that the result holds for $n=1$.  Suppose it holds for $n=N-1$.
Then
\begin{align*}
\prod_{i=1}^N a_i - \prod_{i=1}^N \tilde{a}_i
= \left(\prod_{i=1}^{N-1} a_i\right)(a_N - \tilde{a}_N)
+ \tilde{a}_N \left(\prod_{i=1}^{N-1} a_i - \prod_{i=1}^{N-1} \tilde{a}_i\right).
\end{align*}
Substituting in (\ref{eqn:diffprod}) for $n=N-1$ into the second term, one then easily obtains the result for $n=N$,
proving the inductive step.
\finishproof}

\begin{theorem}
For the first term in (\ref{eqn:twoterms}), the asymptotic limit of the integral and the integral of the asymptotic limit differ
by a rapidly decreasing function:
\begin{align}
\label{eqn:firstterm}
\int \rho_1 u_I \mu e^{\Sprop} = \int \rho_1 u_I \mu e^{\lambda \tilde{S}} + o(\lambda^{-N}) \quad \text{for all }N .
\end{align}
\end{theorem}
{\startproof

Within the support of $\rho_1$, for each $a<b$,
\begin{align*}
e^{\lambda \tilde{S}^\Pi_{ab}} = (x_{ab} + y_{ab})^{2\lambda k_{ab}}
\end{align*}
so that
\begin{align*}
e^{\lambda \tilde{S}^\Pi_{ab}} - e^{S^\Pi_{ab}}
= (x_{ab} + y_{ab})^{2\lambda k_{ab}} - \sum_{m>0}^{\lambda k_{ab}} \binom{2\lambda k_{ab}} {\lambda k_{ab} - m}y_{ab}^{\lambda k_{ab}-m}
x_{ab}^{\lambda k_{ab} + m} .
\end{align*}
Let us first consider the case where $\lambda k_{ab}$ is an integer.  The above then becomes
\begin{align*}
e^{\lambda \tilde{S}^\Pi_{ab}} - e^{S^\Pi_{ab}}
= (x_{ab} + y_{ab})^{2\lambda k_{ab}} - \sum_{i=0}^{\lambda k_{ab}-1} { 2\lambda k_{ab} \choose i }
y_{ab}^{\mathstrut i} x_{ab}^{2 \lambda k_{ab} -i}
= \sum_{i=\lambda k_{ab}}^{2\lambda k_{ab}} { 2\lambda k_{ab} \choose i }y_{ab}^{\mathstrut i}
x_{ab}^{2 \lambda k_{ab} -i}.
\end{align*}
Thus,
\begin{align}
\nonumber
\left|e^{\lambda \tilde{S}^\Pi_{ab}} - e^{S^\Pi_{ab}}\right|
&\le \sum_{i=\lambda k_{ab}}^{2\lambda k_{ab}} { 2\lambda k_{ab} \choose i }|y_{ab}|^{i}
\le { 2\lambda k_{ab} \choose \lambda k_{ab} } \sum_{i=\lambda k_{ab}}^{2\lambda k_{ab}} |y_{ab}|^{i}
= { 2\lambda k_{ab} \choose \lambda k_{ab} } |y_{ab}|^{\lambda k_{ab}}
\sum_{i=0}^{\lambda k_{ab}} |y_{ab}|^i \\
\label{eqn:SPibound}
&= { 2\lambda k_{ab} \choose \lambda k_{ab} } |y_{ab}|^{\lambda k_{ab}}
\frac{1-|y_{ab}|^{\lambda k_{ab} +1}}{1-|y_{ab}|}
\le \frac{32}{25}{ 2\lambda k_{ab} \choose \lambda k_{ab} } |y_{ab}|^{\lambda k_{ab}}
\le \frac{32}{25}{ 2\lambda k_{ab} \choose \lambda k_{ab} } \left(\frac{7}{32}\right)^{\lambda k_{ab}}
\end{align}
where $|x_{ab}|\le 1$ and ${2\lambda k_{ab} \choose i} \le {2\lambda k_{ab} \choose k_{ab}}$
were used in the first and second steps, and the fact that $|y_{ab}|\le 7/32$ within the support of
$\rho_1$ was used in the last two steps.
Let $\tilde{C} > 1$ be arbitrarily chosen. From Stirling's estimate, there exists $\lambda_o$ such that $\lambda > \lambda_o$ implies
\begin{align*}
{ 2\lambda k_{ab} \choose \lambda k_{ab} } < \tilde{C} \frac{2^{2 \lambda k_{ab}}}{\sqrt{\pi \lambda k_{ab}}} .
\end{align*}
(\ref{eqn:SPibound}) thus becomes
\begin{align}
\label{eqn:projbound}
\left|e^{\lambda \tilde{S}^\Pi_{ab}} - e^{S^\Pi_{ab}}\right|
< C \frac{(28/32)^{\lambda k_{ab}}}{\sqrt{\pi \lambda k_{ab}}}
\end{align}
for $C = (32/25) \tilde{C}$.
The case of $\lambda k_{ab}$ half integer can be handled through similar means, yielding
exactly the same bound (\ref{eqn:projbound}) for all $\lambda > \lambda_o'$ for some $\lambda_o'$
independent of the integration variables.
%
%
Thus,
\begin{align*}
\left|\int \rho_1 u_I \mu e^{\Sprop} - \int \rho_1 u_I \mu e^{\lambda \tilde{S}}\right|
&= \left|\int \rho_1 u_I \mu e^{\lambda S^{\rm EPRL}}\left(\prod_{a<b} e^{S^\Pi_{ab}} -
\prod_{a<b} e^{\lambda \tilde{S}^\Pi_{ab}}\right)\right| \\
&\le \int \rho_1 u_I |\mu| \left|\prod_{a<b} e^{S^\Pi_{ab}} -
\prod_{a<b} e^{\lambda \tilde{S}^\Pi_{ab}}\right|
\le \sum_{a<b} \int \rho_1 u_I |\mu| \left|e^{S^\Pi_{ab}} - e^{\lambda \tilde{S}^\Pi_{ab}}\right| \\
&
\le \left(C \sum_{a<b} \int \rho_1 u_I |\mu| \right) \frac{(28/32)^{\lambda k_{ab}}}{\sqrt{\pi \lambda k_{ab}}}
\end{align*}
for all $\lambda > \max\{\lambda_o, \lambda_o'\}$ and
where lemma \ref{lem:diffprod} together with $|e^{S^\Pi_{ab}}|\le 1$ and
$|e^{\lambda \tilde{S}^\Pi_{ab}}|\le 1$ was used in the penultimate step.
Since the right hand side is rapidly decreasing, the result follows.
\finishproof}

\begin{corollary}
\label{cor:first}
The first term in equation (\ref{eqn:twoterms}) is asymptotic to precisely the asymptotics
of the proper vertex amplitude given in section \ref{sec:main_result}.
\end{corollary}
{\startproof

This follows by applying the extended stationary phase theorem \cite{hormander1983}
to the right hand side of (\ref{eqn:firstterm}), and using lemma \ref{lem:criteq}, together with the fact
that the Hessian of matrix of $\Sprop$ is equal to $\lambda$ times the Hessian matrix of $\tilde{S}$
at critical points, a fact shown in \cite{cev2015}.
%
%
\finishproof}

\subsection{Suppression of the second term}
\label{sec:secondterm}

\subsubsection{Partition of unity}

We define $\varphi_{F,\lambda}:\tilde{M} \rightarrow [0,1]$ by
\begin{align*}
\varphi_{F,\lambda} := \varphi\left(\frac{1}{\lambda}\Re S^\Pi + 1\right).
\end{align*}
%
%
so that
$\varphi_{F,\lambda} = 1$ in a neighborhood of the maximal points of $\Re S^\Pi$ and
$\varphi_{F,\lambda} = 0$ in a neighborhood of each point where $\Re S^\Pi$
approaches minus infinity. More specifically, if $\varphi_{F,\lambda} \neq 1$, then
$\Re S^\Pi <  -\frac{1}{4}\lambda$, and if $\varphi_{F,\lambda} \neq 0$, then
$\Re S^\Pi >  -\frac{3}{4}\lambda$.
Furthermore, because $\varphi_{F,\lambda} = 0$ where $\Re S^\Pi$ is non-smooth,
$\varphi_{F,\lambda}$ is smooth throughout $\tilde{M}$.
The integral in the second term of (\ref{eqn:twoterms})
becomes\footnote{Note
that in this integral, the integration over the degenerate sector has been dropped.  This doesn't matter
because it is a set of measure zero.
%
%
Strictly speaking, in fact, the degenerate sector was never included in the integration
because $\Sprop$ is not defined there.}
%
%
%

\begin{align}
\label{eqn:inttwo}
\int \!\!\! \rho_2 u_I \mu e^{\Sprop(\lambda)}
= \int \!\!\! \left(1-\varphi_{F,\lambda}\right) \rho_2 u_I \mu e^{\Sprop(\lambda)} + \int \!\!\! \varphi_{F,\lambda} \rho_2 u_I \mu e^{\Sprop(\lambda)}.
\end{align}

\subsubsection{Suppression}

We next address the two terms in \eqref{eqn:inttwo} in turn, showing that they are sub-dominant.
\begin{description}
\setlength{\parindent}{1em}
\item[The $(1-\varphi_{F,\lambda})$ term]

In this region $\Re S^\Pi < -\frac{1}{4}\lambda$. Defining $\mu' := \left(1-\varphi_{F,\lambda}\right) \rho_2 u_I \mu$ and noting that $\mu'$ is bounded by construction (denote the bound by $M$), we can write:
\begin{align}
\Big| \int \!\!\! \left(1-\varphi_{F,\lambda}\right) \rho_2 u_I \mu e^{\Sprop(\lambda)} \Big| \le \int |\mu'| e^{\Re \Sprop} \le \int M e^{-\frac{1}{4}\lambda} \le A e^{-\frac{1}{4}\lambda}
\end{align}
for some positive constants $M$ and $A$, where in the last step we used the compactness of the integration domain (here $A$ is a positive constant). Therefore, we conclude that this term is $o(\lambda^{-N})$ for all $N$.

\item[The $\varphi_{F, \lambda}$ term]

In the following, we assume that this term is similar to the term evaluated in \ref{sec:first},
in that the asymptotic limit of the integrand of this term in \eqref{eqn:inttwo}
commutes with its integral.  That is, we assume
\begin{align*}
\int \varphi_{F,\lambda} \rho_2 u_I \mu e^{\Sprop(\lambda)} \sim \int  \tilde{B}(\lambda) \varphi_{F,\lambda} \rho_2 u_I \mu \tilde{\mu} e^{\lambda \tilde{S}}.
\end{align*}
$B(\lambda)$ is constant in $2^{10}$ different regions $\mathcal{R}_J$ of $\tilde{M}$.
Expressing the right hand side above as a sum over those regions, the above becomes
\begin{align}
\label{manyints}
\int \varphi_{F,\lambda} \rho_2 u_I \mu e^{\Sprop(\lambda)} \sim C \sum_{J=1}^{2^{10}} \lambda^{-n_J/2}
\int_{\mathcal{R}_J} \varphi_{F,\lambda} \rho_2 u_I \mu \tilde{\mu} e^{\lambda \tilde{S}}
\end{align}
where $C$ and $n_J \ge 0$ are constants independent of $\lambda$.
Referring to \eqref{eqn:SPabt} we note
\begin{align}
\Re \tilde{S}^\Pi_{ab}:= \left\{
\begin{array}{ll}
2 k_{ab} \log |x_{ab}+y_{ab}| & \text{ if }|x_{ab}|>|y_{ab}|\text{ and }|x_{ab}+y_{ab}|^2\ge|4x_{ab}y_{ab}| \\
k_{ab} \log |4 x_{ab} y_{ab}| & \text{ if }|x_{ab}|\le|y_{ab}|\text{ or }|x_{ab}+y_{ab}|^2<|4x_{ab}y_{ab}|
\end{array}
\right.
\end{align}

We now show that $\Re \tilde{S}_{ab}$ is bounded within the support of $\varphi_{F,\lambda}$, 
which is needed in our application of stationary phase in what follows.
Suppose $\Re \tilde{S}^\Pi_{ab} \rightarrow \infty$ at some point.
\begin{itemize}
\item
If $\Re \tilde{S}^\Pi_{ab}=2 k_{ab} \log |x_{ab}+y_{ab}|$, this implies that $|x_{ab}+y_{ab}| \rightarrow 0$, which together with $|x_{ab}+y_{ab}|^2\ge|4x_{ab}y_{ab}|$ in turn implies that $x_{ab} \rightarrow 0$.
\item
If $\Re \tilde{S}^\Pi_{ab}=k_{ab} \log |4 x_{ab} y_{ab}|$, then we have $|4 x_{ab} y_{ab}| \rightarrow 0$ meaning that either $x_{ab} \rightarrow 0$ or $y_{ab} \rightarrow 0$ (or both). However, it is not possible for $y_{ab} \rightarrow 0$ while $x_{ab}$ stays finite because this form of $\Re \tilde{S}^\Pi_{ab}$ only holds when either $|x_{ab}|\le|y_{ab}|$ or $|x_{ab}+y_{ab}|^2<|4x_{ab}y_{ab}|$. In both cases, $y_{ab}$ approaching zero implies that $x_{ab} \rightarrow 0$, so that 
$x_{ab} \rightarrow 0$. 
\end{itemize}
We therefore see that $\Re \tilde{S}^\Pi_{ab} \rightarrow -\infty$ implies
$x_{ab} \rightarrow 0$, which by \eqref{eqn:Ainxy} implies that $\Re S^\Pi_{ab} \rightarrow -\infty \,\, \forall \lambda$. By contraposition it follows that 
if $\exists \lambda \,\,\text{such that} \,\, \Re S^\Pi_{ab}$ has an infimum in a given region, 
then $\Re \tilde{S}^\Pi_{ab}$ has an infimum in that same region.
In fact, in the support of $\varphi_{F, \lambda}$, $\Re S^\Pi > -\frac{3}{4}\lambda$, so that by the above argument $\Re \tilde{S}^\Pi$ has an infimum and hence is bounded.

At each point in the degenerate sector, one or more of the normals 
$n_{ab}[X]$, and hence $\tilde{S}^\Pi$,
is ill-defined with no continuous extension.  
However, not all of the critical point equations of $\tilde{S}$ depend on 
$\tilde{S}^\Pi$: In particular $\Re \tilde{S}^{EPRL}$ and $\delta_z \tilde{S} = 0$
do not depend on $\tilde{S}^\Pi$ and hence are independent of the normals 
$n_{ab}[X]$ and so \textit{are} well-defined and continuous throughout $M$,
including at the degenerate sector.  These critical point equations then select
a submanifold of $M$, which we hereafter refer to as a \textit{critical surface}.
The rest of the critical point equations \textit{do} depend on $n_{ab}[X]$ and so 
are well-defined at a given point of the degenerate sector only when approached from
a fixed direction. As a consequence, these remaining critical point equations select a 
space of \textit{directions} of approach to each point in the degenerate
sector, which we hereafter refer to as the space of \textit{critical directions} at
each such point. As shown in the appendix \ref{app:gen_stat}, the asymptotic behavior of the critical point contributions in this case is affected by the interplay between the restriction on the set of directions of approach and the dimension of the critical surface. 

To apply the result in \ref{app:gen_stat}, we begin by evaluating the dimension of the critical surface in the degenerate sector. The maximality condition $\Re S^{EPRL} = 0$ and the condition of $\delta_z S = 0$ yield three equations:
\begin{align}
\label{eqn:maxeprl}
\xi_{ab} \hateq X_a^{-1} z_{ab}, \quad
 & J\hat{\eta}_{ba} \hateq  X_b^{-1} z_{ab}, \\
\label{eqn:zvar}
\text{and}\quad\; (X^{\dagger}_a)^{-1} \xi_{ab} \hateq & \, (X^{\dagger}_b)^{-1} J\eta_{ba}
\end{align}
As a first step, we consider the critical surface $\mathcal{N}_0$ in $M$ defined by these three equations. The dimension of this critical surface will be denoted by $n_0$. 
The dimension of $M$ is $6\times 4 + 2\times 10 + 2 \times 10 = 64$. 
For each of the 10 pairs $a,b$ with $a<b$, (\ref{eqn:maxeprl}) and (\ref{eqn:zvar}) yield 6 real constraints, 
\textit{unless} $X_{ab} \in \SU(2)$, in which case the last two constraints (\ref{eqn:zvar}) are redundant, leaving only 
4 constraints.
Because of this, the critical surface naturally divides into branches in which certain of the $X_{ab}$ are in $SU(2)$
or not, and the counting will be different for each of these branches.  We consider one such branch at a time.
Note that when $X_{ab} \in \SU(2)$ the two tetrahedra $a$ and $b$ are in the same frame. Because of this,
each branch of the critical surface $\mathcal{N}_0$ can be labeled by a partition of the set of the tetrahedra $\{0, 1, 2, 3, 4\}$. The order of each set in the partition corresponds to the number of tetrahedra in the same frame. The partitions, represented by the order of their elements, are given in the table below (singleton sets are omitted).

When $N$ tetrahedra are in the same frame we have a reduction in dimension of $3(N-1)$ from the fact that corresponding $X_{ab}$'s are in $\SU(2)$. At the same time, compared to the non-degenerate case, $2\binom{N}{2}$ ($\binom{n}{k}$ is a binomial coefficient) restrictions are removed because the above two equations are equivalent. 
Therefore, the total dimension of $\mathcal{N}_0$ in a branch defined by a given partition is 
$64 - 60+2\sum_i \binom{N_i}{2}-3\sum_i (N_i-1)$, where sums are performed over elements of the partition,
and $N_i$ is the order of the corresponding element.
\begin{center}
  \begin{tabular}{ l | r }
    \hline    
    Partition & Dimension \\ \hline
    2 & 3 \\
    3 & 4\\ 
    4 & 7\\
    5 & 12\\
    (2,2) & 2\\
    (2,3) & 3\\
    \hline
  \end{tabular}
\end{center}
From this table it follows that the dimension $n_0$ of the critical surface associated with the equations \eqref{eqn:maxeprl} and \eqref{eqn:zvar} falls between $2$ and $12$.

Now we use another critical point condition, namely the maximality of $\Re \tilde{S}^{\Pi}$. Recalling that 
\begin{equation}
\label{eq:spisum}
 \tilde{S}^{\Pi} = \sum_{a<b} \tilde{S}^{\Pi}_{ab} 
\end{equation}
we have two possibilities for each of the $10$ pairs $(ab)$:
\begin{enumerate}
\item 
\begin{align}
\label{eq:logx+y}
\tilde{S}^\Pi_{ab} = 2k_{ab} \log(x_{ab}+y_{ab}) = 2k_{ab} \log(\langle \hat{\eta}_{ba}, \xi_{ba} \rangle).
\end{align}
Maximality then implies that $ \hat{\eta}_{ba}=\xi_{ba}$, which reduces the dimensionality of the critical surface by $2$.
\item 
\begin{align}
\label{eq:log4xy}
\tilde{S}^{\Pi}_{ab} = k_{ab}\log(4x_{ab}y_{ab}).
\end{align}
From appendix \ref{app:max}, 
maximality then implies :  $|\scal{\hat{\eta}_{ba},\nu_{ba}}| = |\scal{\nu_{ba},\xi_{ba}}| = \frac{1}{\sqrt{2}}$. Note that $\hat{\eta}_{ba} \neq \xi_{ba}$ for this form of $\tilde{S}^{\Pi}_{ab}$. It follows that in this case the dimension of
the space of critical \textit{directions} is reduced by $2$.
\end{enumerate}

Let $q_1$ be the number of terms in \eqref{eq:spisum}  of the form \eqref{eq:logx+y} and $q_2$ be the number of terms of the form \eqref{eq:log4xy}, so that $q_1+q_2=10$. If $q_1=10$, then there are no critical points in the degenerate sector and the asymptotics are exponentially suppressed.  
To see that there are no critical points in this case, note that maximality in this case implies $[\eta_{ba}] = [\xi_{ba}]$, which together with (\ref{maxcrit}), (\ref{crit2}), and the closure of $(k_{ab}, \xi_{ab}, \xi_{ba})$ imply that $(X_a, z_{ab}, \xi_{ab}, \xi_{ba})$ satisfy the same critical point equations as in \asymptoticL. But by assumption the boundary data $(k_{ab}, \xi_{ab},\xi_{ba})$ is compatible with 
a non-degenerate Lorentzian 4- simplex, so that the only critical points are those with $(X_a)$ non-degenerate
\asymptoticL, which are explicitly excluded from the support of $\varphi_{F,\lambda}$.

If not all terms in \eqref{eq:spisum} have the form \eqref{eq:logx+y} (i.e. $q_1<10$), the dimension $n$ of the critical surface is bounded from above by $n_0-2q_1$ if $2q_1<n_0$ and $0$ otherwise. The reduction $s$ in the number of critical directions  is determined by the number of terms of the form \eqref{eq:log4xy} as $s = 2q_2$. Hence we derive that for $q_1<10$:
\[ s-n \geq  \begin{cases}
        20-n_0  & \text{when } n_0>2q_1 \\
        2q_2 & \text{otherwise}
        \end{cases}
\]
This shows that $s-n>0$, and applying the theorem in the appendix \ref{app:gen_stat}, we can conclude  that the contributions from the critical surface in the degenerate sector will be sub-dominant by a factor of $\lambda^{-\frac{s-n}{2}}$ in the asymptotics. Additional critical point equations will further reduce the number of critical directions (increasing $s$) or the dimension of the critical surface (decreasing $n$), thus making this term even more sub-dominant.
Lastly, the extra $\lambda^{-n_J/2}$ factor present in each term of (\ref{manyints}) 
makes the sub-dominance even stronger.

\end{description}

\section{Conclusion}

In this paper, we have studied the asymptotics of the Lorentzian proper vertex amplitude introduced in
\cite{ez2015}, in the semi-classical limit of large spins. Since the path integral expression for the amplitude contains
a new projector action term which is only asymptotically linear in spins and possesses new discontinuities,
its asymptotic analysis presented a technical challenge. New techniques were developed to generalize
stationary phase methods for the path integral in question.
In order to apply these techniques, we deduced the asymptotic form of the action.
Critical points were found and their contributions were evaluated.
Partitions of unity were introduced in order to handle separately contributions from
regions of the integration manifold containing discontinuities.
In particular, the degenerate sector where at least one $X_{ab}$ is in $\SU(2)$ was considered separately.

We have shown that the Lorentzian proper vertex amplitude has the correct semi-classical limit for non-degenerate boundary
data,
so that the single Feynman term corresponding to the Einstein-Hilbert sector is isolated.
Thus it avoids the mixing of different Plebanski sectors and the extra term present in the asymptotics of the EPRL amplitude is eliminated in this case.
Furthermore, the coefficient of the one remaining term is the same as the coefficient of the corresponding
term in the EPRL asymptotics, as is shown in the companion work \cite{cev2015},
a work which also calculates the graviton propagator from the present vertex.

Let us close with a final remark. Many complications in the above analysis can be avoided by choosing a stronger version of the projector. Specifically, if one replaces \eqref{eqn:quantum_projector}
by
\begin{align*}
\Pi_{ba}(\{X_{a'b'}\}):=&\Pi_{((k_{ab}-1)\|\mathbf{v}_{ba}[\{X_{a'b'}\}]\|,\infty)}\left(\mathbf{v}_{ba}[\{X_{a'b'}\}]_i\,
\rotoperator^i \right)\\
&=\Pi_{((k_{ab}-1),\infty)}\left(\frac{\mathbf{v}_{ba}[\{X_{a'b'}\}]}{\|\mathbf{v}_{ba}[\{X_{a'b'}\}]\|}\,
\rotoperator^i \right)
\end{align*}
where ${\bf v}_{ba}[\{X_{a'b'}\}]:=\beta_{ba}(\{\widehat{X}_{a'b'}\})\;\tr(\pauli_i\, X_{ba}\,X_{ba}^{\dagger})$ 
then $S^{\Pi}_{ba}$ reduces to $2 k_{ab} \log x_{ab}$. Thus, the modified action scales linearly in $k_{ab}$ and one does not need to consider two asymptotic forms as in \eqref{eqn:SPabt} 
However, such a strong modification is physically questionable as it inserts by hand a classical equation of motion 
--- namely $\mathbf{v}_{ba}[\{X_{a'b'}\}] \propto \mathbf{n}_{\xi_{ba}}$ ---
into the quantum theory, and thus may kill physical quantum fluctuations. 


{\acknowledgments
%
%
We thank Benjamin Bahr, Christopher Beetle, Atousa Chaharsough Shirazi, Muxin Han and Wojciech Kaminski  for discussions.
In particular, we thank Wojciech Kaminski for help in the application of the dominated convergence theorem in section
\ref{sec:projasym}.
J.E. and I.V. were supported in part by NSF grants PHY-1205968 and PHY-1505490
and J.E. by NASA through the University of Central Florida's NASA-Florida Space Grant Consortium.
A.Z. acknowledges financial support
by the grant of Polish Narodowe Centrum Nauki
 nr 2012/05/E/ST2/03308.
}

\appendix

\section{Consequence of maximality of one of the asymptotic projector actions}
\label{app:max}

In this appendix, 
we note an interesting implication of the maximality condition $\Re \tilde{S}^\Pi_{ab} = 0$
in the case when $\Re \tilde{S}^\Pi_{ab} = k_{ab} \log |4x_{ab}y_{ab}|$.  
This implication is used in section \ref{sec:secondterm} of the main text. 
From $\Re \tilde{S}^\Pi_{ab}=0$ it follows that $|4x_{ab}y_{ab}|=1$. Using the definitions of $x_{ab}, y_{ab}$ this becomes:
\begin{align*}
4 |\scal{\hat{\eta}_{ba},\nu_{ba}} \scal{\nu_{ba},\xi_{ba}}|\,|\scal{\hat{\eta}_{ba},J\nu_{ba}} \scal{J\nu_{ba},\xi_{ba}}| = 1.
\end{align*} 
Since $\hat{\eta}_{ba}$ is normalized, it implies that
\begin{align*}
|\scal{\hat{\eta}_{ba},J\nu_{ba}}| = \sqrt{1-|\scal{\hat{\eta}_{ba},\nu_{ba}}|^2}
\end{align*}
and similarly for $|\scal{J\nu_{ba},\xi_{ba}}|$. Therefore, we have
\begin{align*}
4 |\scal{\hat{\eta}_{ba},\nu_{ba}} \scal{\nu_{ba},\xi_{ba}}| \sqrt{1-|\scal{\hat{\eta}_{ba},\nu_{ba}}|^2} \sqrt{1-|\scal{\nu_{ba},\xi_{ba}}|^2} = 1.
\end{align*}
Define a function $f(a) := a \sqrt{1-a^2}$ for $a \in \R^{+,0}$. Then the previous equation becomes
\begin{align*}
4 f(|\scal{\hat{\eta}_{ba},\nu_{ba}}|) f(|\scal{\nu_{ba},\xi_{ba}}|) = 1.
\end{align*}
Noting that the maximum of $f(a)$ is $\frac{1}{2}$ and it is attained only when $a=\frac{1}{\sqrt{2}}$ we conclude that the above equation implies that
\begin{align}
\label{eqn:max4xynu}
|\scal{\hat{\eta}_{ba},\nu_{ba}}| = |\scal{\nu_{ba},\xi_{ba}}| = \frac{1}{\sqrt{2}}.
\end{align}

\section{Stationary phase contributions from critical surfaces and critical directions}
\label{app:gen_stat}

In this appendix, we extend the usual stationary phase argument in two ways (1.) to the case of a \textit{submanifold} of
stationary points, and (2.) to the case where the critical point equations are satisfied only when approached from certain directions.  The latter is of course only possible when the action in the integral is not smooth at the point or points in question.

Consider an $m$-dimensional compact manifold $\mathcal{M}$, and an $n$-dimensional submanifold $\mathcal{N}$.
Fix a vector space $V_p  < T_p\mathcal{M}$
%
%
at each point of $p$ of $\mathcal{N}$ such that $T_p\mathcal{N} \subsetneq V_p$ (so that $V_p$ contains at least one
direction transverse to $\mathcal{N}$).
Assume the dimension of $V_p$ is independent of $p$ and denote this dimension by $m-s$,
so that $n <  m-s$, which implies
\begin{align}
\label{snrelation}
s < m-n.
\end{align}
We now consider an integral
\begin{align}
\label{Idef}
I(\lambda) := \int_\mathcal{M} e^{\lambda S} d\mu
\end{align}
with action $S: \mathcal{M} \rightarrow \C$ and measure $d\mu$. Let us assume that $\nabla S$ and $\Re S$ both vanish
\textit{at each point $p$ of $\mathcal{N}$ when approached from any direction within $V_p$.}  We refer to $\mathcal{N}$
as a `critical surface' and $V_p$ as a space of `critical directions.' 

The question the present appendix answers is: In the asymptotic limit of large $\lambda$, what does the presence of this critical surface with the given critical directions contribute to the integral $I(\lambda)$, and is this contribution dominant or subdominant as compared to what would be contributed by an isolated bulk critical point.

In answering this question, we make a few assumptions.
First, in order to focus on the contribution from the critical surface $\mathcal{N}$,
we assume there are no critical points or critical directions of $S$ other than those
just describe. 
We furthermore make the usual assumptions used in stationary phase theorems \cite{hormander1983}, appropriately adapted to include the case of a critical surface with 
critical directions. Namely, that
\begin{enumerate}
\item Both $\mu$ and $S$ are smooth in $\mathcal{M}\setminus \mathcal{N}$ and 
are bounded in $\mathcal{M}$.
\item Along any direction of approach to any point of $\mathcal{N}$, 
$\mu$ has continuous first order derivative, and $S$ has continuous third order derivative,
\item \label{negasump} ${\rm Re} S$ is everywhere non-positive, 
\item \label{degasump} In the limit as one approaches any point of $\mathcal{N}$ from any transverse direction, 
the kernel of the real part of the Hessian of $S$ is exactly equal to the tangent space
of $\mathcal{N}$.
\end{enumerate}
%
%

By breaking up the integral (\ref{Idef}) with partitions of unity in the usual way, we may reduce the integral
to smaller integrals for which global coordinate charts exist. For this reason, we may without loss of generality 
assume that a global coordinate chart exists. 
We choose this chart $(\x{0}, \dots, \x{m-1})$ 
to be adapted to $\mathcal{N}$ and the vector spaces $V_p$ in the sense that 
\begin{enumerate}
\item $\mathcal{N}$ is the $n$-dimensional submanifold defined by the equations $\x{0} = \x{1} = \dots = \x{m-n-1} = 0$, so
that the tangent space to $\mathcal{N}$ at each point is spanned by
$\frac{\partial}{\partial \x{m-n}}, \dots, \frac{\partial}{\partial \x{m-1}}$.
\item At each point $p \in \mathcal{N}$, $V_p$ is spanned by
$\frac{\partial}{\partial \x{s}}, \dots,  \frac{\partial}{\partial \x{m-1}}$
\end{enumerate}
%
%
%
Define $\mu$ by $d\mu = \mu \,  d^m x$.  The integral (\ref{Idef}) then becomes
\begin{align}
\label{Ix}
I(\lambda) := \int e^{\lambda S} \mu \, d^m x .
\end{align}

Next, we introduce generalized spherical coordinates for $\x{0}, \dots, \x{s}$.
For this, we first extend the range of integration for these coordinates as needed so that
it becomes a ball in $\R^{s+1}$, without extending the support of $\mu$. 
The new coordinates are then defined by 
\begin{equation}
\label{genspherical}
\begin{aligned}
\x{i} &= \y{s} \left(\prod_{j=0}^{i-1} \cos \y{j} \right) \sin \y{i} \quad &&\text{ for } i = 0, \dots s-1 \\
\x{s} &= \y{s} \left(\prod_{j=0}^{s-1} \cos \y{j} \right)
\end{aligned}
\end{equation}
where each product symbol is defined to be one when the upper limit is less than the lower limit, and 
where $y_i$ takes values in $[-\pi/2, \pi/2]$ for $i=0, \dots , s-2$,
$y_{s-1}$ takes values in $[-3\pi/2, \pi/2]$, and $y_s>0$.\footnote{These angles are related to the more
usual angular coordinates $\phi_i$ via $y_i = \pi/2 - \phi_i$.  This choice has
been made so that $\x{i} = 0$ if $y_i = 0$ for $i=0, \dots, s-1$.
}
The volume element for the first $s+1$ $x$ coordinates in terms of the above new coordinates is then
\begin{align}
\label{sphvol}
d^{s+1}x = (y_s)^s (\sin^{s-1} y_0) (\sin^{s-2} y_1) \cdots (\sin y_{s-1}) d^{s+1}y .
\end{align}
Define
\begin{align}
\label{ydef}
\y{i} &:= \x{i} \text{ for }i=s+1, \dots m-1
\end{align}
With (\ref{sphvol}) and (\ref{ydef}), expression (\ref{Ix}) can then be written
\begin{align}
\label{Iy}
I(\lambda) := \int e^{\lambda S} \mu  \cdot (\y{s})^s (\sin^{s-1} y_0) (\sin^{s-2} y_1) \cdots (\sin \y{s-1}) d^m y
\end{align}

In terms of the new coordinates $\y{i}$, the critical points and directions take a much simpler form, 
as we prove in the following
\begin{lemma}
\label{ycritlemm}
Suppose $\gamma(t)$ is a curve in $\mathcal{M}$.
Then 
\begin{align}
\label{critcond}
\lim_{t \rightarrow 0} \frac{\partial S}{\partial x^i}(\gamma(t)) = 0 \text{ for all } i \text{ and } 
\lim_{t \rightarrow 0} \Re S(\gamma(t)) = 0
\end{align} 
 if and only if
$\lim_{t\rightarrow 0}\y{i}(\gamma(t)) = 0$ for $i=0, \dots m-n-1$.
\end{lemma}
{\startproof

The derivative of equations (\ref{genspherical}) and (\ref{ydef}) along the curve $\gamma(t)$ yield:
\begin{alignat}{2}
\dot{\x{i}} &= \dot{\y{s}} \left(\prod_{j=0}^{i-1} \cos \y{j} \right) \sin \y{i}
+  \y{s} \frac{d}{dt}\left[\left(\prod_{j=0}^{i-1} \cos \y{j} \right) \sin \y{i} \right]
&&\quad \text{ for } i = 0, \dots s-1 \\
\label{xyderiv}
\dot{\x{s}} &= \dot{\y{s}} \left(\prod_{j=0}^{s-1} \cos \y{j} \right) +
\y{s} \frac{d}{dt}\left(\prod_{j=0}^{s-1} \cos \y{j} \right) && \\
\dot{\x{i}} &= \dot{\y{i}} \quad &&\text{ for } i=s+1, \dots m-1
\end{alignat}

($\Leftarrow$)

Suppose 
$\lim_{t \rightarrow 0}\y{i}(\gamma(t)) = 0$ for $i=0, \dots, m-n-1$.
(\ref{genspherical}) then implies
$\lim_{t \rightarrow 0} \x{i}(\gamma(t)) = \x{i}(\gamma(0)) = 0$ for $i=0, \dots, m-n-1$, so that $\gamma(0) \in \mathcal{N}$,
and from (\ref{xyderiv}), $\dot{\x{i}} = 0$ for $i=0, \dots, s-1$, so that $\dot{\gamma}(0) \in V_{\gamma(0)}$, 
a critical direction.  Thus (\ref{critcond}) follows.

($\Rightarrow$)

Suppose (\ref{critcond}) holds, so that 
$\gamma(0) \in \mathcal{N}$ and $\dot{\gamma}(0) \in V_{\gamma(0)}$.
One then has
\begin{align}
\label{Ncond}
\x{i}(\gamma(0)) &= 0 \quad \text{ for }i=0, \dots, m-n-1, \text{ and }\\
\label{Vcond}
\dot{\x{i}}(\gamma(0)) &= 0 \quad \text{ for }i=0, \dots, s-1.
\end{align}
From (\ref{snrelation}), $s \le m-n-1$, so that (\ref{Ncond}) gives
\begin{align*}
\lim_{t \rightarrow 0}\y{s}(\gamma(t)) = \sqrt{\sum_{i=0}^s (\x{i}(\gamma(0))^2} = 0
\end{align*}
where the continuity of (\ref{Ncond}) is used.
The $i=0$ case of (\ref{Vcond}) then gives $\lim_{t \rightarrow 0}y_0(\gamma(t)) = 0$, afterwhich 
the $i=1$ case of (\ref{Vcond}) gives $\lim_{t\rightarrow 0}y_1(\gamma(t)) = 0$, etc.,
until one has shown in the end that
$\lim_{t\rightarrow 0}\y{i} = 0$ for $i=0, \dots s-1$. 
 Lastly, for $i \in [s+1, m-n-1]$, (\ref{Ncond}) trivially implies $\lim_{t\rightarrow 0}\y{i}(\gamma(t))=0$.
\finishproof }

It follows that,  in the limit of large $\lambda$, the only part of the integral (\ref{Iy}) that is not rapidly decreasing 
is that in an arbitrarily small neighborhood of $\y{i}=0$ for $i<m-n$.
To make this precise, for each $\epsilon > 0$, let $\mathcal{M}_{\epsilon}$
denote the region of $\mathcal{M}$ in which the coordinates $\y{i}$
are restricted to $\y{i}\in (-\epsilon, \epsilon)$ for $i<m-n$. 
We then have
\begin{align}
\label{Iepsilon}
I(\lambda) \sim \int_{\mathcal{M}_\epsilon} 
e^{\lambda S} \mu  \cdot (\y{s})^s (\sin^{s-1} y_0) (\sin^{s-2} y_1) \cdots (\sin \y{s-1}) d^m y
\end{align}
for each $\epsilon>0$. 

Consider now Taylor expansions of $\mu$ and $S$ in the coordinates 
$\y{i}$ for $i<m-n$. Specifically, define
\begin{align*}
\tilde{S}(y) &= S(0, \y{\ge m-n})
 + \sum_{i<m-n}\frac{\partial S}{\partial y_i}(0, \y{\ge m-n}) y_i  
 + \frac{1}{2}  \sum_{i,j<m-n}\frac{\partial^2 S}{\partial y_i \partial y_j}(0, \y{\ge m-n}) y_i y_j \\
\mu_0(y) &= \mu(0, \y{\ge m-n}) .
\end{align*}
%
Taylor's theorem then gives
\begin{align*}
S(y) &= \tilde{S}(y) + o\left((\y{<m-n})^2\right) \\
\mu(y) &= \mu_0(y) + o\left((\y{<m-n})^0\right) .
\end{align*} 
If we define
\begin{align}
\label{Jdef}
J_\epsilon(\lambda) := \int_{\mathcal{M}_\epsilon} e^{\lambda \tilde{S}} \mu_0  \cdot (\y{s})^s (\sin^{s-1} y_0) (\sin^{s-2} y_1) \cdots (\sin \y{s-1}) d^m y,
\end{align}
using $m-n\ge 1$, equation (\ref{Iepsilon}) becomes
\begin{align}
\label{Jasym}
I(\lambda) \sim J_\epsilon(\lambda)+ o(\epsilon).
\end{align}
From now on we assume that the $o(\epsilon)$ term above, which can be made arbitrarily small, can be neglected.
Note that the determinant of the Jacobian for the transformation from $\y{i}$ to $\x{i}$ appears in (\ref{sphvol}):
\begin{align*}
\det\left(\frac{\partial \x{i}}{\partial \y{j}}\right) =  r^s (\sin^{s-1} \alpha_0) (\sin^{s-2} \alpha_1) \cdots (\sin \alpha_{s-1})
\end{align*}
which is bounded.  This fact, together with lemma \ref{ycritlemm} and the chain rule, implies that the linear term
in $\tilde{S}(y)$ vanishes, and the constant term is pure imaginary.  Thus, if we define 
$B(\y{\ge m-n}):= \Im S(0, \y{\ge m-n})$ and 
$\A{ij}(\y{\ge m-n}):= \frac{1}{2}  \sum_{i,j<m-n}\frac{\partial^2 S}{\partial y_i \partial y_j}(0, \y{\ge m-n})$,
$\tilde{S}(y)$ becomes
\begin{align*}
\tilde{S}(\lambda) = i B(\y{\ge m-n})
 + \sum_{i,j<m-n}\A{ij}(\y{\ge m-n})) y_i y_j .
\end{align*}

We now turn to bounding the integral $J_\epsilon(\lambda)$. 
Using the triangle inequality, and $|\sin y_i| \le 1$, (\ref{Iy}) yields
\begin{align}
\nonumber
|J_\epsilon(\lambda)| \le \int e^{\lambda \Re \tilde{S}} \mu_0  \cdot (\y{s})^s  d^m y
& = \int e^{\lambda \sum_{i,j <m-n} \Re \A{ij} \y{i} \y{j} } \mu_0 \cdot  (\y{s})^s  d^m y \\
\label{Jbound}
& = \int e^{\lambda \sum_{i,j < m-n} \Re \A{ij} \y{i} \y{j} } \mu_0  \cdot (\y{s})^s  d^m y 
\end{align}
$\Re A$ is a symmetric bilinear form, and 
by assumptions \ref{negasump} and \ref{degasump} about $S$ at the start of this appendix, is also negative definite. 
Let $\vbasis{i}{j}$, $i=0, \dots, m-n-1$ be any basis of $\R^{m-n}$  which
is `orthonomal' with respect to $A$ in the sense that $\sum_{i,j<m-n}\A{kl}\vbasis{i}{k} \vbasis{j}{l} = - \covdelta{ij}$.
Define new coordinates $\ucoord{i}$, $i=0, \dots m-n-1$ by $y^i =: \sum_{j<m-n}\ucoord{j} \vbasis{j}{i}$, so that (\ref{Jbound}) becomes
\begin{align}
\label{uint}
|J_\epsilon(\lambda)| & \le
\int  m(\y{\le m-n}) \left(\sum_{i<m-n}\ucoord{i} \vbasis{i}{s}\right)^s  
\exp\left(-\lambda \sum_{i<m-n}  (\ucoord{i})^2\right) d^{m-n}u \cdot d^n y
\end{align}
where $m(\y{m-n})$ takes values in $\R^+$ and is bounded.
To isolate the $\lambda$ dependent part of the integral we perform one more change of variables to spherical coordinates
\begin{align*}
\ucoord{i} &= \rho \left(\prod_{j=0}^{i-1} \cos \beta_j \right) \sin \beta_j \quad \text{ for }i=0, \dots m-n-2, \quad \text{ and}\\
\ucoord{m-n-1} &= \rho \left(\prod_{j=0}^{m-n-2} \cos \beta_j \right) 
\end{align*}
The range of the $\beta_i$'s is the same as the $\alpha_i$'s above, and the Jacobian is of the same form (\ref{sphvol}).
Noting (1.) that the measure factor in (\ref{uint}) depends on $\rho$ by an overall factor of $\rho^{s}$, (2.)
that all $\lambda$ dependence can be pulled out of the $\beta_i$ integrals, and (3.) that the range of the integration of the $\beta_i$ integrals are independent of $\rho$, we see that evaluation of the $\beta_i$ integrals and 
remaining $y_i$ integrals in (\ref{uint}) gives
\begin{align}
\label{rhoint}
|I(\lambda)| \le \text{(const.)} \int_0^a \rho^{m-n-1+s} e^{-\lambda \rho^2} d\rho
\end{align}
where $a$ is the supremum of the values that $\rho$ takes in the support of the integral.
To evaluate the above integral, we need the following lemma.
\begin{lemma}
for any $a > 0$ and any non-negative integer $q$, 
\begin{align}
\label{asymptint}
\int_{0}^a \rho^q e^{-\lambda \rho^2} d\rho \sim \text{(const.)} \lambda^{-\frac{1}{2}(1+q)}
\end{align}
asymptotically for large $\lambda$.
\end{lemma}
{\startproof

For the case $q=0$ we have 
\begin{align*}
\int_{0}^a e^{-\lambda \rho^2} d\rho = \frac{1}{\sqrt{\lambda}} \int_{0}^{a\sqrt{\lambda}}e^{-w^2} dw
= \frac{1}{\sqrt{\lambda}}\frac{\sqrt{\pi}}{2}\left( {\rm erf}(a\sqrt{\lambda})\right)
\sim \frac{\sqrt{\pi}}{2} \lambda^{-\frac{1}{2}},
\end{align*}
matching the right hand side in this case.  Here ${\rm erf}$ denotes the error function and we have used that the error function
is asymptotic to $1$.
The case $q=1$ can be calculated explicitly using the change of variables $w:=u^2$.
To obtain the rest of cases, note that if $f \sim g$ and $\lim f = \lim g = 0$, then by L'H\^opital's rule
$f' \sim g'$.  Applying this fact to the cases $q=0$ and $q=1$ repeatedly then yields the rest of the cases.

\finishproof}

Application of this lemma to (\ref{rhoint}) yields
\begin{align*}
J_\epsilon(\lambda) = o\left(\lambda^{-\frac{1}{2}(m-n+ s)}\right) .
\end{align*}
Thus (\ref{Jasym}) becomes
\begin{align*}
I(\lambda) \sim o\left(\lambda^{-\frac{1}{2}(m-n+ s)}\right) 
\end{align*}
which is equivalent to
\begin{align*}
I(\lambda) = o\left(\lambda^{-\frac{1}{2}(m-n+ s)}\right) .
\end{align*}

%
%

%

\end{document}